\documentclass[12pt]{article}
\usepackage{amsmath}
\input{epsf.sty}

\numberwithin{equation}{section}

\newcommand{\nn}{\nonumber}
\newcommand{\vs}[1]{\vspace*{#1}}

\newcommand{\p}{\partial}
\newcommand{\Half}{\frac12}
\newcommand{\unit}{\hbox to 3.8pt{\hskip1.3pt \vrule height 7.4pt
    width .4pt \hskip.7pt \vrule height 7.85pt width .4pt \kern-2.4pt 
    \hrulefill \kern-3pt \raise 3.7pt\hbox{\char'40}}}

\def\href#1#2{#2}

\begin{document}


\begin{titlepage}

\title{
\hfill\parbox{4cm}{
{\normalsize UT-Komaba/02-02}\\[-5mm]
{\normalsize\tt hep-th/0204203}
}
\\[20pt]
Dynamical Decay of  \\
Brane--Antibrane and Dielectric Brane
}
\author{
Koji {\sc Hashimoto}
\\[10pt]
{\it Institute of Physics, University of Tokyo,}\\
{\it Komaba, Meguro-ku, Tokyo 153-8902, Japan}\\
{\small E-mail: {\tt koji@hep1.c.u-tokyo.ac.jp}}\\
}
\date{\normalsize April, 2002}    

\maketitle
\thispagestyle{empty}

\begin{abstract}
\normalsize\noindent
Using D-brane effective field theories, we study dynamical
decay of unstable brane systems : (i) a parallel 
brane-antibrane pair with separation $l$ and (ii) a dielectric brane.
In particular 
we give explicitly the decay width of these unstable systems,
and describe how the decay proceeds after the tunnel effect. 
The decay (i) is analysed by the use of a tachyon
effective action on the D$p$-$\overline{\mbox{D}p}$. 
A pair annihilation starts by nucleation of a bubble of a tachyon domain 
wall which represents a throat connecting these branes, and
the tunneling decay width $\Gamma$ is found to be proportional to 
$\sim\exp (-l^{p+1} T_{{\rm D}p})$.
We study also the decay leaving topological defects corresponding to
lower-dimensional branes, which may be relevant for recent inflationary 
braneworld scenario. 
As for the decay (ii), first we observe that D$p$-branes generically 
``curl up'' in a nontrivial RR field strength. Using this viewpoint, we
compute the decay width of the dielectric D2-branes by constructing
relevant Euclidean bounce solutions in the shape of a funnel. 
We also give new
solutions in doughnut shape which are involved with nucleation of 
dielectric branes from nothing.

\end{abstract}

\end{titlepage}

\tableofcontents

\clearpage

\section{Introduction}

Recently, the importance of unstable brane systems has been recognised
widely. One of the most successful utilization of the unstable brane
systems was Sen's conjectures \cite{Sentac1} which enable one to study
the classification of the D-branes by K-theories and to define new open  
string field theories, via  tachyon condensation.  However, most of the
tachyon condensation considered in the literatures have been static. The
next step toward obtaining information on the determination of the true 
vacuum of string/M theory may be to analyse and understand the 
time-dependent dynamics of the decay of the unstable brane systems.

The main obstacles to progress along this important direction is
that string theory contains infinite number of field degrees of freedom.
The recent observation of ``rolling
tachyon'' by A.\ Sen \cite{Senroll} has picked up an open string 
tachyon field in unstable brane systems, and studied 
how the homogeneous tachyon rolls down the potential. 
There the initial conditions of the tachyon field became
important. However, it might be plausible 
to consider the situation where the
decay starts from a false vacuum, and a quantum tunneling
provides a classically unstable initial condition, then the unstable
mode rolls down the potential hill.\footnote{See \cite{Felder} 
for recent progress on the spontaneous symmetry breaking called tachyon
preheating. }

In this paper, we investigate two typical unstable systems which are
classically stable but cause a decay through the tunnel effect: 
(i) a D$p$-brane and an anti-D$p$-brane located in parallel with 
separation $l$, and (ii) dielectric branes \cite{myers}
in constant Ramond-Ramond (RR) field 
strength. These brane systems are nonperturbatively unstable and decay
through the quantum tunneling. As for
(i), the parallel branes are expected to nucleate a throat bridge which
connects these two branes by the tunnel effect, and then this throat
expands to sweep out the brane-antibrane pair. Concerning (ii), 
the dielectric brane (which is classically stable) 
instantly expands to form a larger spherical D-brane which is
classically unstable,
and then the large spherical brane starts expanding gradually
due to its classical instability.  
The situations (i) and (ii) are similar to each other
as for the processes of the decay, although the description of these
branes are completely different. The situation (i) is described by the
condensation of the tachyon field coming from open strings connecting
the brane and the antibrane, while the situation (ii) is described by
the Dirac-Born-Infeld (DBI) 
action of the dielectric brane in the constant RR
field strength. However, we will find their similarity in the analysis
of the decay. 

This paper is organised as follows. The paper consists of 
two parts which are almost independent of each other.
The first part (Sec.\ 2) treats the decay of the dielectric branes, and
the second part (Sec.\ 3) deals with the decay of the 
brane-antibrane. The second part uses the implication from Sec.\ 2.
Each part has introductory remarks. 

In Sec.\ 2, we observe that the
branes in the background RR field strength are generically curled up,
and utilizing this coordinate choice for the brane worldvolume we
compute the 
decay width of the dielectric branes. The relevant bounce configuration
in the Euclidean target space is found to be a funnel-shaped. In
addition, another new doughnut-shaped branes are found to describe the
bounce solution for nucleating a dielectric brane from nothing. 

In Sec.\ 3, first we construct an action of the tachyon field theory on
the brane-antibrane separated by $l$. Then we obtain the bounce solution
by following Coleman's method to give the decay width. By checking
the scalar field configuration on this tachyon bounce background we find
an evidence for that the bounce solution is actually describing a throat
connecting the brane and the antibrane. 
We apply this method also to find the width of the decay with
leaving lower dimensional stable D-branes as topological defects. This
may be relevant for the recent inflationary braneworld scenario. 
We also consider the D0-brane anti-D0-brane decay by combining and
generalizing the results of Sec.\ 2.

We conclude with several remarks and comments on the future directions
in the final section.
The  appendices treat
two topics: how the tachyon potential found in Sec.\ 2 is corrected
to the higher order in the framework of the boundary string field theory
(BSFT), and how the introduction of the non-commutativity induces stable 
non-commutative solitons on the brane-antibrane.


\section{Decay of dielectric brane}

\subsection{Introductory remarks}

It is well known that D-branes and strings can decay via the tunnel
effect in the background of the constant R-R field strength 
\cite{ext,Emparan,Park}. Along similar argument, one can find that 
the Myers' dielectric brane can decay via the tunnel effect. 
The final aim of this section is to compute the decay rate of the
dielectric branes for a given background and a given
number of constituent D0-branes in the dielectric brane. 

To see that the dielectric branes can decay, 
let us briefly review what was discussed by R.\ C.\ 
Myers in \cite{myers}. 
The effective action describing a D$p$-brane in the flat target space
metric with a nontrivial RR gauge field background is\footnote{We are
following the notation of \cite{myers}.}
\begin{eqnarray}
 S = T_{{\rm D}p} \int dt d^{p}x 
\left[
-\sqrt{-\det(\eta_{\mu\nu} + \p_\mu X^i \p_\nu X^i 
+ 2\pi{\alpha'} F_{\mu\nu})} 
+ C(X)
\right]
\label{Dac}
\end{eqnarray}
where the worldvolume directions are specified by
$\mu, \nu = 0,1,\cdots,p$ and we have introduced the transverse scalar
fields $X^i$ with $i=p+1,\cdots,9$. Considering especially a D2-brane
and adopting a constant RR field strength 
\begin{eqnarray}
 [dC]_{tijk} = 
\left\{
\begin{array}{ll}
-2f \epsilon_{ijk} & \mbox{for}\;\; i,j,k \in \{1,2,3\} \\
0& \mbox{otherwise}
\end{array}
\right.
\label{d2b}
\end{eqnarray}
we observe a stable spherical D2-brane configuration in the following
way. Since the background is symmetric in the directions $x^1, x^2,
x^3$, we may put a spherical symmetry ansatz for the worldvolume. 
Then the brane configuration is described only by the radius $r$. 
Turning on the magnetic field on the brane appropriately, one finds the
energy of the spherical brane as
\begin{eqnarray}
 E = T_{{\rm D}2}
\left[
A(S^2) \sqrt{r^4 + \pi^2 {\alpha'}^2 N_{\rm D0}^2}-
2f V(S^2)r^3
\right],
\label{myene}
\end{eqnarray}
where $A(S^p)$ and $V(S^p)$ are the area of $S^p$ with the unit radius 
and the volume of the region enclosed by that $S^p$, respectively. 
$N_{\rm D0}$ is the number of
the D0-branes bounded in the spherical 
D2-brane, coming from the worldvolume gauge field. 
Myers found that by expanding
the square root there exists a local minimum of the potential energy. 
One concludes that in the constant RR 4-form field strength $N_{\rm D0}$
D0-branes can expand and form a spherical bound state with a local
D2-brane charge.

\begin{figure}[t]
\begin{center}
\begin{minipage}{80mm}
\begin{center}
   \leavevmode
   \epsfxsize=60mm
\put(-5,110){$E/4\pi T_{{\rm D}2}$} 
\put(160,5){$r$} 
\put(35,58){$\times$} 
\put(35,49){A} 
\put(115,59){$\times$} 
\put(130,59){B} 
\put(80,59){$\longrightarrow$} 
\put(90,96){$\times$} 
\put(90,105){A'} 
  \epsfbox{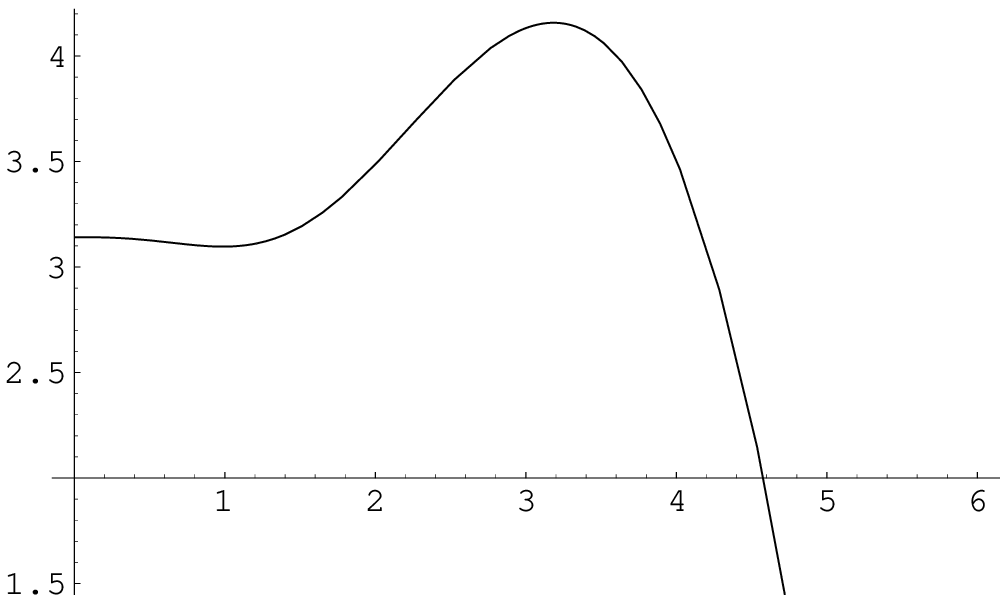}
   \caption{Potential energy for the spherical D2-brane with the
 magnetic flux on it. We chose $N_{{\rm D0}}=1$ in the unit 
${\alpha'}=1$. The point A is a local minimum which corresponds to the
dielectric brane. It decays by quantum tunneling to the larger spherical
brane (B). }
   \label{figa}
\end{center}
\end{minipage}
\hspace{5mm}
\begin{minipage}{70mm}
\begin{center}
   \leavevmode
   \epsfxsize=60mm
\put(-5,110){$E/4\pi T_{{\rm D}2}$} 
\put(160,50){$r$} 
\put(93,97){$\times$} 
\put(93,85){C} 
   \epsfbox{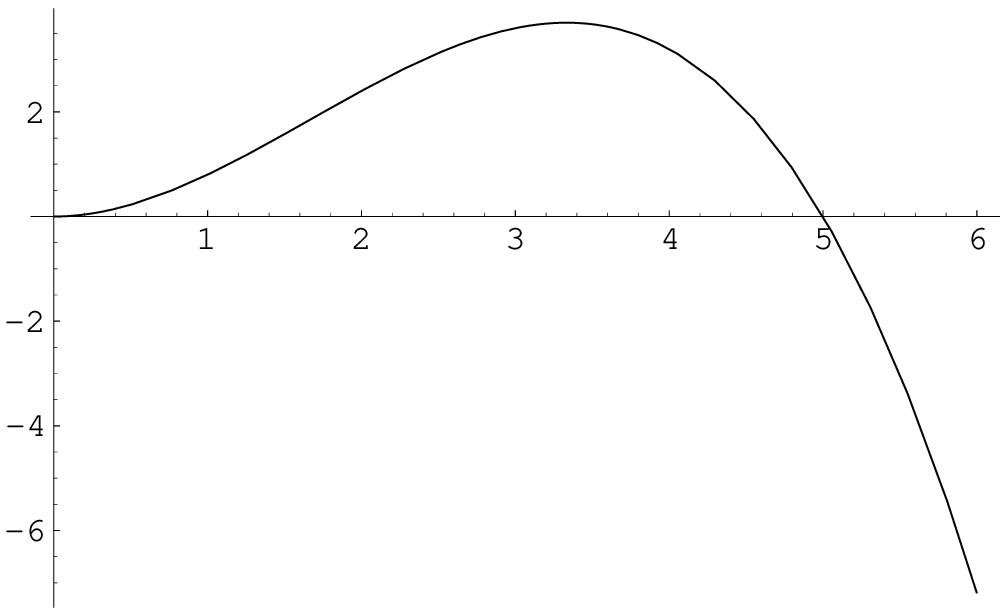}
   \caption{Potential energy with no magnetic flux. There is no local
 minimum except $R=0$.} 
   \label{figb}
\end{center}
\end{minipage}
\end{center}
\end{figure}

One can get the exact radius of the dielectric 
brane by finding the extrema of the potential energy (\ref{myene}), 
\begin{eqnarray}
 r_M= \sqrt{
\frac{1}{2f^2}-\sqrt{\frac{1}{4f^4}- \pi^2 {\alpha'}^2 N_{\rm D0}^2}
}.
\label{myr}
\end{eqnarray}
The energy (\ref{myene}) is plotted in Fig.\ \ref{figa}. 
One can easily find that this radius of the dielectric branes 
is merely a local minimum, and thus the dielectric branes can decay via
the tunnel effect to form a larger sphere which is classically 
unstable and thus expands. Another extremum is given by the radius
\begin{eqnarray}
 r_{\rm top}= \sqrt{
\frac{1}{2f^2}+\sqrt{\frac{1}{4f^4}- \pi^2 {\alpha'}^2 N_{{\rm D0}}^2}
}.
\label{rtop}
\end{eqnarray}
This radius corresponds to the top of the hill in Fig.\ \ref{figa} 
(the point
A'). The potential height at that point may characterize the decay
width of the dielectric D2-brane. 

We understand that this tunnel effect occurs also with no initial
D0-brane : For $N_{\rm D0}=0$, 
the local minimum is simply $r=0$, thus nothing. It can
decay by producing a large spherical D2-brane. The energy potential is 
shown in Fig.\ 2. This is a generalization of Schwinger process
\cite{Schwinger}, 
nucleation of an electron-positron pair in a constant electric
field.\footnote{Nucleation of a monopole-antimonopole pair was studied
in \cite{monop}, and a generalization to a spherical membrane can be
found in \cite{cole3}. }
This phenomenon is not limited to the
D2-brane case. For any constant RR background $F^{(p+2)}$, a spherical 
D$p$-brane formation occurs, and it expands so as to decrease the energy
of the background RR-form. The width of the decay by this spherical
brane formation was obtained in \cite{teit}. If the target spacetime
dimension is $p+2$, the spherical brane sweeps out the RR background,
although in this paper we will not consider this kind of back reaction. 

In the following subsections, we introduce a description (a choice of
the worldvolume parametrization) which is useful to describe D$p$-branes
in the relevant background RR fields. This new description enables us to
compute the decay width of the dielectric branes. 


\subsection{Curled-up branes and reformulation of dielectric effect}
The constant background RR field strength which we consider in this
section is, for a given $p$, 
\begin{eqnarray}
 [dC]_{ti_1\cdots i_{p+1}} = 
\left\{
\begin{array}{ll}
-2f \epsilon_{i_1\cdots i_{p+1}} & \mbox{for}\;\; i_n 
\in \{1,2,\cdots, p+1\} \\
0& \mbox{otherwise}
\end{array}
\right..
\label{backp}
\end{eqnarray}
This is solved in a certain gauge as
\begin{eqnarray}
 C_{012\cdots p} = 2f X^{p+1}.
\end{eqnarray}
In this background, D$p$-branes are expected to be polarized to form a
sphere $S^p$. However, 
we shall not assume the spherical symmetry of the brane configuration in
the target space. Instead, the worldvolume gauge choice 
\begin{eqnarray}
 t=T, \quad  x^1=X^1,\quad  \cdots,\quad  x^p = X^p
\end{eqnarray}
which is usual will be more useful in the following.

We turn on only a single scalar field $X\equiv X^{p+1}$. 
Then the D-brane action (\ref{Dac}) in the background (\ref{backp})
becomes
\begin{eqnarray}
S &=& T_{{\rm D}p}
\int dt d^{p}x
\left[
-\sqrt{-\det (\eta_{\mu\nu} + \p_\mu X \p_\nu X)}
+2f X\right].
\end{eqnarray}
Here we turn off the worldvolume gauge field for simplicity.
One can see that for the constant RR field strength we have the linear
gauge potential for the Chern-Simons term of the action.

\begin{figure}[tdp]
\begin{center}
\begin{minipage}{70mm}
\begin{center}
   \leavevmode
   \epsfxsize=50mm
  \epsfbox{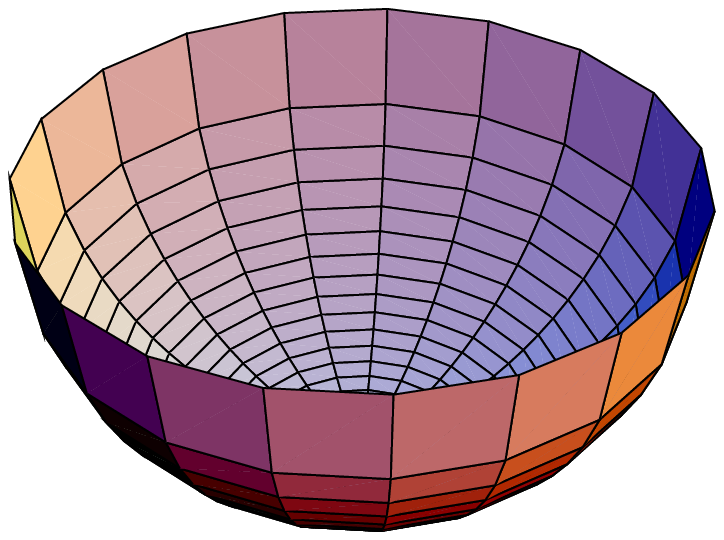}
   \caption{A curled-up brane in the background constant RR field
 strength with negative $f$. }
   \label{figC-1}
\end{center}
\end{minipage}
\hspace{5mm}
\begin{minipage}{70mm}
\begin{center}
   \leavevmode
   \epsfxsize=50mm
   \epsfbox{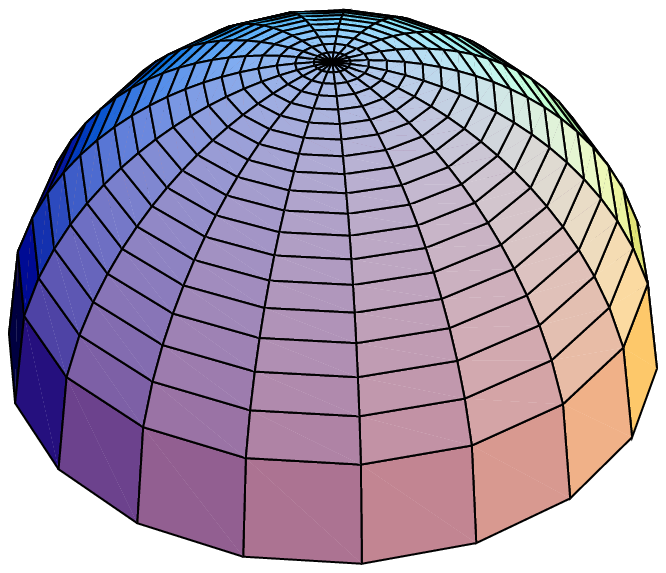}
   \caption{A curled-up brane with positive $f$. }
   \label{figC-2}
\end{center}
\end{minipage}
\end{center}
\end{figure}

Assuming that the scalar field depends only on the distance from the
worldvolume origin, 
$r\equiv \sqrt{(x^1)^2 + \cdots + (x^p)^2}$,
the static action is written as
\begin{eqnarray}
S =  - T_{{\rm D}p}\int dt 
A(S^p)\int r^{p-1} dr 
\left[
\sqrt{1+ (\p_r X)^2 } -2f  X\right].
\end{eqnarray}
Then the equations of motion reads
\begin{eqnarray}
 2fr^{p-1} + \p_r 
\left(\frac{r^{p-1} \p_r X}{\sqrt{1 + (\p_r X)^2}}\right)=0.
\end{eqnarray}
This can be integrated easily and one finds
\begin{eqnarray}
\frac{\p_r X}{\sqrt{1 + (\p_r X)^2}}= -\frac{2fr}{p} 
+ \frac{{c_1}}{r^{p-1}} 
\end{eqnarray}
where ${c_1}$ is an integration constant. This parameter ${c_1}$ will be
important for evaluating the decay width of the dielectric branes. In
this and the next subsections we put ${c_1}=0$ since we need no scalar
charge. For vanishing ${c_1}$, we obtain
\begin{eqnarray}
 \p_r X = \frac{-\displaystyle
\frac{2f}{p}r}{\sqrt{1-\displaystyle\frac{4f^2}{p^2}r^2}}.
\label{soln}
\end{eqnarray}
Immediately it is found that there is a critical radius 
$r=r_0\equiv|p/2f|$ at which $\p_r X$ diverges. 
Therefore the solution is defined only interior the disk whose radius is
$r_0$. Integrating the equation (\ref{soln}), we obtain 
a brane configuration
\begin{eqnarray}
 X = \int^r_{r_0} \!dr \;\p_r X = 
\left\{
\begin{array}{ll}
\sqrt{r_0^2-r^2} & \mbox{for}\;\; f>0, \\
-\sqrt{r_0^2-r^2} & \mbox{for}\;\; f<0. 
\end{array}
\right.
\end{eqnarray}
This is precisely a hemisphere, as shown in Fig.\ \ref{figC-1} and Fig.\  
\ref{figC-2}. 
The D-brane is ``curled up''
due to the constant RR field strength background.
This hemisphere is to be understood as a part of a complete sphere.
To form a complete sphere, we
need both solutions of $f>0$ and $f<0$, to join them together. The
reason why the sign of the RR field flips is that in our worldvolume
coordinate choice the orientation of the lower hemisphere is opposite to
that of the upper hemisphere. This is a kind of a pair of a brane and an
antibrane. This joining is somewhat similar to what has been found in
the decay of the brane-antibrane pair in \cite{Callan, Savvidy} in which
two volcano-shaped branes are combined to form a ``throat'' connecting
a brane and an antibrane.

The result is of course consistent with the configuration obtained with
the spherical ansatz in the previous subsection. In Fig.\ \ref{figb},
there exists a sphere solution at the top of the potential hill (the
point C in the figure).
For $p=2$, the critical radius $r_0$ agrees with the one from the
spherical ansatz (\ref{rtop}) with $N_{{\rm D0}}=0$. 

How the branes are curled up depends on the background RR field
strength. In the present case we have considered only the constant
background, thus we have obtained the sphere which has a constant
curvature of the worldvolume. This curvature depends on the value of the
background RR field, thus for varying background some nontrivial shapes
of the curled up branes are possible. We use this possibility to
describe pair creation of D0-branes in the nontrivial RR background in
Sec.\ 3.7, by explicitly constructing a non-trivially curled-up
Euclidean worldline. 

Let us generalize the above result to the dielectric branes on which
the nontrivial magnetic flux is turned on. Consider a D2-brane action 
with the magnetic field in the constant RR field strength (\ref{d2b}), 
\begin{eqnarray}
S &=& -T_{{\rm D}2}\int dt dx^1 dx^2\left[
\sqrt{-\det (\eta_{\mu\nu} + \p_\mu X \p_\nu X + 2\pi{\alpha'}
F_{\mu\nu})}
-2f X\right] \nn\\
&=& 
-T_{{\rm D}2}\int dt dx^1 dx^2\left[
\sqrt{1+ (\p_\mu X)^2 + 4\pi^2 {\alpha'}^2F_{12}^2} -2f X\right].
\end{eqnarray}
We have already ignored the time dependence.
Assuming the radially symmetric configuration with 
$r\equiv \sqrt{(x^1)^2 + (x^2)^2}$, the equations of motion becomes
\begin{eqnarray}
2fr + \p_r 
\left(\frac{r \p_r X}{\sqrt{1 + (\p_r X)^2+ 4\pi^2 {\alpha'}^2
F_{12}^2}}\right)=0,
\quad
\p_r\left(\frac{2\pi {\alpha'} F_{12}}
{\sqrt{1 + (\p_r X)^2+ 4\pi^2 {\alpha'}^2F_{12}^2}}\right)=0.
\end{eqnarray}
It is easy to integrate these equations as in the same manner. 
the result is 
\begin{eqnarray}
\frac{\p_r X}{\sqrt{1 + (\p_r X)^2 + 4\pi^2 {\alpha'}^2 F_{12}^2}}
= -fr + \frac{{c_1}}{r},
\quad 
\frac{2\pi{\alpha'}F_{12}}{\sqrt{1 + (\p_r X)^2+ 4\pi^2 {\alpha'}^2 
F_{12}^2}}={c_2}.
\label{ccp}
\end{eqnarray}
Here we have two integration constants ${c_1}$, ${c_2}$. 
Again, we simply put ${c_1}=0$ 
since the resultant configuration given below is 
consistent with the boundary condition (finiteness of the whole
dielectric brane configuration). 
The second equation in (\ref{ccp}) 
can be solved for $F_{12}$, giving a relation
between $X$ and $F_{12}$ as 
\begin{eqnarray}
4\pi^2 {\alpha'}^2
 F_{12}^2 = \frac{{c_2}^2}{1-{c_2}^2}\left(1+(\p_r X)^2\right).
\end{eqnarray}
Substituting this into the first equation in (\ref{ccp}), 
we obtain a (hemi-)spherical configuration $X=\sqrt{r_0^2-r^2}$
whose radius is given by
\begin{eqnarray}
 r_0= \frac{\sqrt{1-{c_2}^2}}{|f|}.
\label{radim}
\end{eqnarray}
Certainly, the vanishing of the magnetic field 
${c_2}=0$ leads back to the previous result. 

The total magnetic flux on the D2-brane is equal to the number of
D0-branes. This induced D0-brane charge is expressed as 
\begin{eqnarray}
\frac{N_{{\rm D0}}}{2}\mu_{{\rm D}0}&=&
\mu_{{\rm D}2} 
\int 2 \pi  r dr  (2\pi{\alpha'} F)\nn\\
&=& 2\pi \mu_{{\rm D}2} 
\int_0^{r_0} r dr \frac{{c_2}}{\sqrt{1-{c_2}^2}}
\sqrt{1+(\p_r X)^2} 
\nn\\
&=&
\mu_{{\rm D}2} \frac{2\pi}{f^2} {c_2} \sqrt{1-{c_2}^2}.
\end{eqnarray}
Here $\mu_{{\rm D}p}$ is the RR-charge of the D$p$-brane which is equal
to the D$p$-brane 
tension $T_{{\rm D}p} = ((2\pi)^p {\alpha'}^{(p+1)/2} g)^{-1}$. 
We have only half of the total charge,
$N_{{\rm D0}}/2$, since we made an integration over the hemisphere.
Therefore we have
\begin{eqnarray}
 N_{{\rm D0}} = \frac{c_2\sqrt{1-{c_2}^2}}{\pi{\alpha'} f^2}.
\end{eqnarray}
For a given background $f$ and a D0-brane charge $N_{\rm D0}$, 
it is easy to solve this relation for ${c_2}$. 
We have two solutions as
\begin{eqnarray}
 {c_2}^2 = \Half \pm \sqrt{\frac14 - \pi^2{\alpha'}^2 N_{{\rm D0}}^2f^4}.
\end{eqnarray}
Both are in the region $0<{c_2}<1$. Substituting this into the radius
formula (\ref{radim}), we see that the larger ${c_2}$ is reproducing the 
radius of the dielectric brane $r_{\rm M}$(\ref{myr}), 
while the smaller one is giving the
radius of the unstable solution $r_{\rm top}$ (\ref{rtop}).
Note that there is an upper 
bound for $|N_{{\rm D0}}|$, for a given background $f$. 
The existence condition for dielectric branes is 
\begin{eqnarray}
 |N_{{\rm D0}}| \leq  \frac{1}{2\pi{\alpha'} f^2}.
\label{econ}
\end{eqnarray}


\subsection{Nucleation and expansion of spherical D$p$-brane}
\label{nuex}

For the preparation for the calculation of the decay width of the
dielectric branes in the next subsection, we develop the technique 
of constructing a bounce solution relevant to the tunnel effect. The
simplest setting is the vacuum decay, that is, nucleation of a spherical
D$p$-brane  from nothing, in the background constant RR field strength. 
The computation of the decay width has been done already fifteen years
ago by C.\ 
Teitelboim \cite{teit}. Here we reformulate his results in terms
of our worldvolume coordinate choice for the later purpose. 
We find that our formulation also gives how the nucleated spherical
brane decays for its expansion. 

According to \cite{coleman1}, the tunneling rate
with the potential like Fig.\ \ref{figa} and Fig.\ \ref{figb}
with a false vacuum is expressed as 
\begin{eqnarray}
\Gamma = c \exp[-S_{\rm E}], 
\label{coled} 
\end{eqnarray}
where $S_{\rm E}$ is the Euclidean action for which the bounce solution
of $S_{\rm E}$ is substituted. 
(The evaluation of the coefficient $c$ can be found in \cite{coleman2}
and we shall not evaluate it in this paper.) 
The bounce solution should be a solution with the highest symmetry of
the Euclidean action, thus it would be invariant under the $p+2$
dimensional spacetime rotation. This indicates that the bounce
configuration of the Euclidean D$p$-brane (whose worldvolume is
Euclidean and $p+1$ dimensional) is $S^{p+1}$. 
Let us study this in our language.

It is obvious that the curled-up branes formulated in the previous
sections can be found in the same manner in the Euclidean space. 
The brane configuration of an Euclidean
D$p$-brane in the constant RR $p+2$-form field strength (\ref{backp}) is
governed by\footnote{The Euclidean action is obtained by the following
redefinition: the world sheet time $t \rightarrow i \tau$, the 
target space time $T \rightarrow i T$, the 
action $S \rightarrow -iS_{\rm E}$. Note that in this redefinition the
RR gauge field transforms as a covariant tensor, 
$C_{t12\cdots p} \rightarrow -i C_{\tau 12 \cdots p}$.}
\begin{eqnarray}
S_{\rm E} = T_{{\rm D}p}\int d\tau d^{p}x
\left[
\sqrt{\det (\delta_{\mu\nu} + \p_\mu T \p_\nu T + 2\pi{\alpha'}
F_{\mu\nu})}
-2f T\right]
\end{eqnarray}
where $T$ is the fluctuation along the Euclidean time direction in the
target spacetime, and we took the worldvolume coordinate along the
spatial directions as before. $\tau$ is the worldvolume Euclidean time.
We have assumed that the other scalar fields are vanishing.

For no worldvolume gauge fields turned on, the situation is completely
same as what have been studied in Sec.\ 2.2. 
The only difference is the
dimension. For the decay with the formation of the spherical D$p$-brane
in the background $p+2$ form constant field strength, the relevant
bounce solution is $S^{p+1}$ while the static 
configuration studied in Sec.\ 2.2 was $S^p$.  

Following Sec.\ 2.2, the Euclidean action
\begin{eqnarray}
 S_{\rm E} =
 T_{{\rm D}p}\int d\tau d^{p}x
\left[
\sqrt{\det (\delta_{\mu\nu} + \p_\mu T \p_\nu T)}
-2f T\right]
\end{eqnarray}
can be solved easily. For simplicity, we assume $f>0$ in the following. 
The solution is  a $p+1$ dimensional hemisphere,
\begin{eqnarray}
T = \sqrt{r_0^2 - r^2},
\quad
r_0 \equiv \frac{p+1}{2f}
\label{solE}
\end{eqnarray}
where $r \equiv \sqrt{\tau^2+(x^1)^2 + (x^2)^2 + \cdots + (x^p)^2}$. 
Then, the action evaluated is the same form as what has been found 
in \cite{teit}: 
\begin{eqnarray}
2 S_{\rm E} = A(S^{p+1})r_0^{p+1} -2f V(S^{p+1}) r_0^{p+2}.
\label{actionde}
\end{eqnarray}
Here the multiplication factor 2 is for the enveloping of two
hemispheres. 
Using a relation $A(S^{p+1}) = (p+2)V(S^{p+1})$, 
we can see that this $r_0$ (\ref{solE}) is what makes the
action (\ref{actionde}) be maximized.  

The effect of our coordinate choice appears in the calculation of the
brane decay after the nucleation. The expansion is due to the potential
in Fig.\ \ref{figb}, since the nucleation radius $(p+1)/2f$ (\ref{solE})
is larger than the radius at the top of the hill $p/2f$. Hence the
configuration rolls down the hill toward the infinite radius. 

The Minkowskian action is given as
\begin{eqnarray}
 S = T_{{\rm D}p}\int d^{p+1}\xi 
\left[-\sqrt{-\det(\eta_{MN}\p_a\Phi^M \p_b \Phi^N)}
+ C(\Phi)\right].
\end{eqnarray}
We introduce a space-like parametrization of the
worldvolume coordinates as before, by choosing a gauge
\begin{eqnarray}
\xi^1 = X^1, \quad 
\cdots, \quad \xi^{p+1} = X^{p+1}.
\label{gaugec}
\end{eqnarray}
This turns out to be useful for solving the equations of motion. 
We switch on only the fluctuation scalar for the target space 
time direction $T$. Then the action is reduced to 
\begin{eqnarray}
 S=T_{{\rm D}p}\int A(S^p) r^pdr
\left[\sqrt{-1+(\p_r T)^2}
-2fT\right].
\end{eqnarray}
The equations of motion can be integrated in the same manner, giving
\begin{eqnarray}
\frac{\p_r T}{\sqrt{-1+(\p_r T)^2}} = 
\frac{2f}{p+1}r + \frac{{c_3}}{r^p}.
\end{eqnarray}
As we will see shortly, the boundary condition at the nucleation time 
can be satisfied with the 
integration constant ${c_3}=0$. Then we have
\begin{eqnarray}
\p_rT = \frac{\displaystyle
\frac{2f}{p+1}r}{\sqrt{-1+\displaystyle\frac{4f^2}{(p+1)^2}r^2}}.
\end{eqnarray}
Thus, appropriately choosing the zero of the time, we have the brane
configuration 
\begin{eqnarray}
T= \sqrt{r^2 - \frac{(p+1)^2}{4f^2}}.
\end{eqnarray}
In the above expression, the radius is
$r = \sqrt{(X^1)^2 + \cdots + (X^{p+1})^2}$, as understood 
from the gauge choice 
(\ref{gaugec}). Therefore this equation shows that the 
shape of the solution is the hyperboloid in the target space.
At $r=r_0=(p+1)/2f$, the initial velocity of the bubble is zero, as
should be 
satisfied from the continuity condition with the nucleation radius
(\ref{solE}). 
The bubble expands with the speed approaching that of light.

\begin{figure}[tdp]
\begin{center}
\begin{minipage}{120mm}
 \begin{center}
   \leavevmode
   \epsfxsize=70mm
\put(0,45){$T=0 \longrightarrow$}
\put(85,140){$T$}
\put(130,30){$\biggm\}$ Euclidean evolution}
\put(180,80){Minkowskian evolution}
   \epsfbox{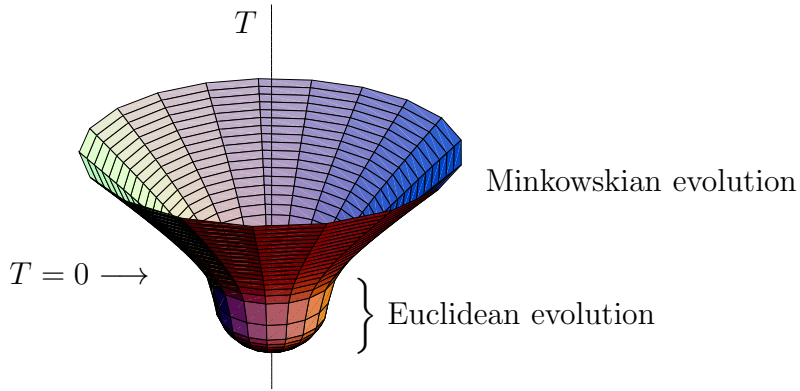}
   \caption{Nucleation of a spherical D-brane and its expansion.
   This figure is especially for the decay of a circular D-string
  ($p=1$) in the background RR 3-form field strength. }
   \label{fige}
  \end{center}
\end{minipage}
\end{center}
\end{figure}

How the decay proceeds
is shown in Fig.\ \ref{fige}. The lower half of the figure is
the nucleation part evolving in the Euclidean time, and the upper half
is the expansion of the brane while the Minkowskian time. The picture is
just like the birth of inflationary universes \cite{Vilenkin}.

\subsection{Decay width of dielectric D2-brane}

Now we are ready to study the decay of a dielectric D2-brane. 
The relevant bounce configuration should satisfy the following
conditions : 
\begin{itemize}
 \item 
The initial boundary condition is a static spherical D2-brane 
with magnetic flux and the radius (\ref{myr}). 
\item
The highest symmetry is SO(3), thus the bounce configuration must be
invariant under this.
\item
The final boundary condition is a spherical D2-brane whose radius is
larger than the radius of the top of the hill, (\ref{rtop}).
\end{itemize}

In the situation of the previous subsection, one needs no initial brane
configuration and no magnetic flux, thus the symmetry was SO(4) and the
brane configuration was a hemisphere. However, this time, $S^2$ of 
the dielectric D2-brane (the initial boundary condition) should be
smoothly connected in the Euclidean spacetime. Hence the total shape of
the solution must be a funnel.\footnote{Although 
this funnel 
looks similar to what has been investigated in \cite{Emparan, Park}
which is 
called ``spherical bulge'', the crucial difference from ours is the
dimensionality and the branes considered.  
In those papers, the decay of a {\it fundamental
string} into a {\it toroidal} D2-brane with 
gauge flux was considered. Due to their assumption of
homogeneous nucleation along the string, their argument has different 
dimensionality. }

Let us start with the Euclidean D2-brane action in the gauge chosen as
before, (\ref{gaugec}).
The action is 
\begin{eqnarray}
 S_{\rm E} = T_{{\rm D}2}\int d^3\xi \left[
\sqrt{\det(\delta_{\mu\nu} + \p_\mu T \p_\nu T + 2\pi{\alpha'}
F_{\mu\nu})} 
-2fT
\right],
\end{eqnarray}
where $\mu, \nu = 1,2,3$, and $T$ is the scalar field representing the
displacement along the Euclidean time direction. 
Turning on the gauge fields, we have
\begin{eqnarray}
 S_{\rm E} = 
T_{{\rm D}2}\int d^3\xi \left[
\sqrt{1+ (\p_iT)^2 + 4 \pi^2 {\alpha'}^2 B_i^2 
+4 \pi^2 {\alpha'}^2 (B_i \p_i T)^2} -2fT
\right].
\end{eqnarray}
Since now the worldvolume is the Euclidean 3 dimensional space, 
the gauge field strength was arranged into a magnetic field.
As mentioned before, the symmetry ansatz adopted here is SO(3), thus 
the Bianchi identity $ \p_i B_i =0$ is solved as 
\begin{eqnarray}
2 \pi {\alpha'} B_i = \p_i \left(\frac{-{c_5}}{r}\right),
\label{solb}
\end{eqnarray}
where $r \equiv\sqrt{(\xi^1)^2+(\xi^2)^2+(\xi^3)^2}$. 
Using this magnetic field configuration, we need only to solve the
equation of motion for $T$:\footnote{The reason why we need not to solve
the gauge field equations of motion is as follows. Let us introduce a 
Lagrange multiplier field $\chi$ which has another
meaning of a magneto-static potential \cite{Gary}, as
\begin{eqnarray}
{\cal L} = f(B_i, \p_i T) 
- B_i \p_i \chi. 
\end{eqnarray}
Then we can regard the magnetic field $B_i$ as an independent field free
from the Bianchi identity. Solving the equations of motion for $B_i$
and substitute it back to the original action, that is, performing the
Legendre transformation, the resultant Lagrangian will be
$\widetilde{\cal L} = g(\p_i T, \p_i \chi)$. However, the equations of
motion for this $\chi$ is just the Bianchi identity
$\p_iB_i=0$. Therefore, since we have solved this under the SO(3)
symmetry ansatz, we only need to solve the equations of motion for $T$. 
}
\begin{eqnarray}
 \p_i \left(
\frac{\p_i T + 4 \pi^2 {\alpha'}^2 B_i (B_j \p_j T)}
{\sqrt{1 + (\p_i T)^2 + 4 \pi^2 {\alpha'}^2 B_i^2 
+4 \pi^2 {\alpha'}^2 (B_i \p_i T)^2}} 
\right)=-2f.
\end{eqnarray}
This can be integrated as
\begin{eqnarray}
\frac{\p_r T + 4 \pi {\alpha'}^2 B_r (B_r \p_r T)}
{\sqrt{1 + (\p_r T)^2 + 4 \pi {\alpha'}^2 B_r^2 + 
4 \pi {\alpha'}^2(B_r \p_r T)^2}} 
=-\frac{2f}{3}r + \frac{{c_4}}{r^2},
\end{eqnarray}
where ${c_4}$ is the integration constant which we shall not put zero
this time. Substituting the solution of $B_i$ (\ref{solb}), we obtain an
expression for the solution $T$ as
\begin{eqnarray}
 \p_r T = \frac{\displaystyle-\frac{2fr}{3} + \frac{{c_4}}{r^2}}
{\sqrt{1-\displaystyle\frac{4f^2 r^2}{9}+\frac{4f {c_4}}{3r}
+ \frac{1}{r^4} \left({c_5}^2 - {c_4}^2\right)}}.
\label{soldeca}
\end{eqnarray}

Before evaluating the action using this solution, 
let us investigate the meaning of the
integration constants appearing in the solution. First, 
the magnetic field characterized by the monopole charge ${c_5}$ is
expected to be related with the D0-brane charge induced on the
dielectric brane. In our case, the dielectric brane is an initial state,
and in the bounce solution above, this must be a fixed Euclidean time
slice $T=T_0$. Because we adopt the rotational symmetry, this Euclidean
time slice should corresponds to a certain fixed radius $r=r_{\rm M}$
in the worldvolume, through the solution (\ref{soldeca}). Therefore, 
the D0-brane charge can be evaluated from the magnetic field as
\begin{eqnarray}
 N_{{\rm D0}} \mu_{{\rm D}0} &=& 
\mu_{{\rm D}2}\int_{r=r_{\rm M}}
2\pi{\alpha'} F_{ij} \Half dx^i \wedge dx^j
=4\pi {c_5}\mu_{{\rm D}2}.
\label{evaf}
\end{eqnarray}
Thus 
\begin{eqnarray}
N_{{\rm D0}}= \frac{1}{\pi{\alpha'}} {c_5}.
\end{eqnarray}

\begin{figure}[tdp]
\begin{center}
\begin{minipage}{90mm}
\begin{center}
   \leavevmode
   \epsfxsize=80mm
  \epsfbox{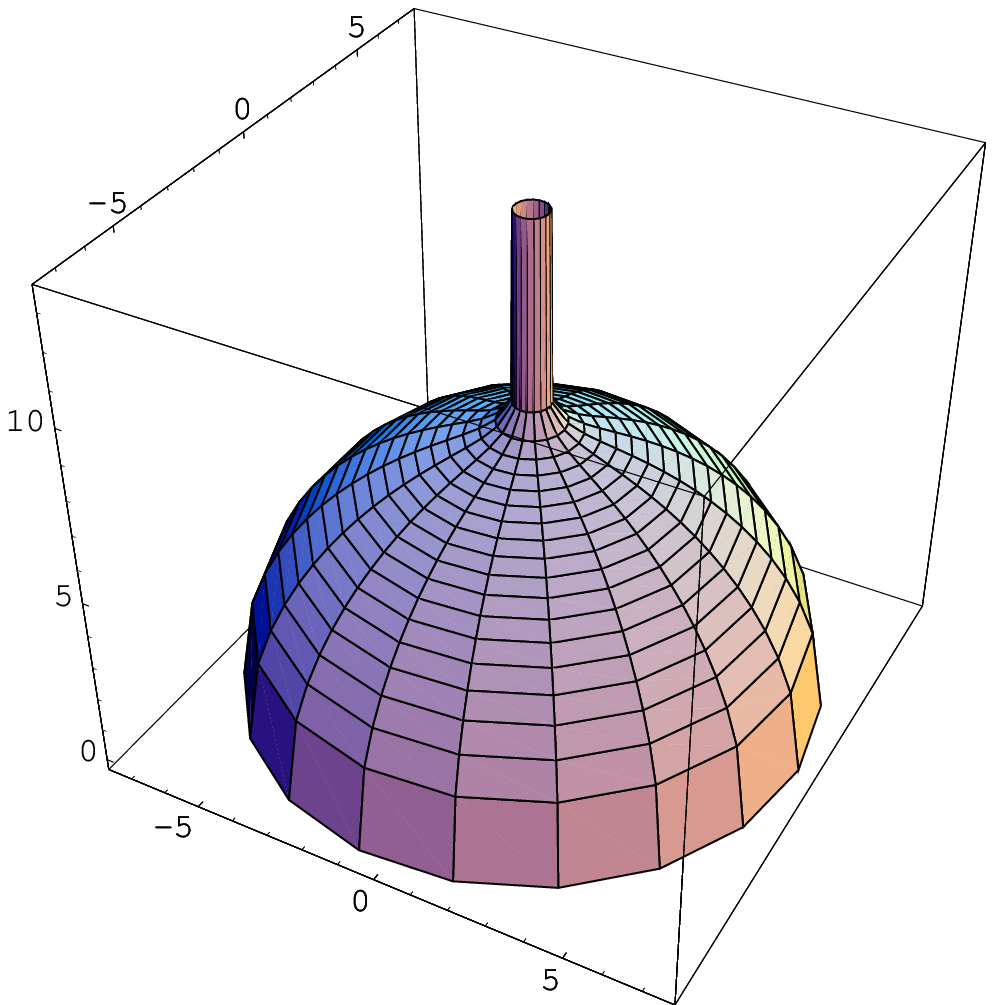}
   \caption{The bounce solution relevant for the decay of the dielectric 
 brane with $N_{\rm D0}=1$ in the background $f=0.2$ in the unit
 $2 \pi {\alpha'}=1$. Note that this figure must be understood as
 upside-down.}  
   \label{figf-1}
\end{center}
\end{minipage}
\hspace{5mm}
\begin{minipage}{50mm}
\begin{center}
   \leavevmode
   \epsfxsize=30mm
   \epsfbox{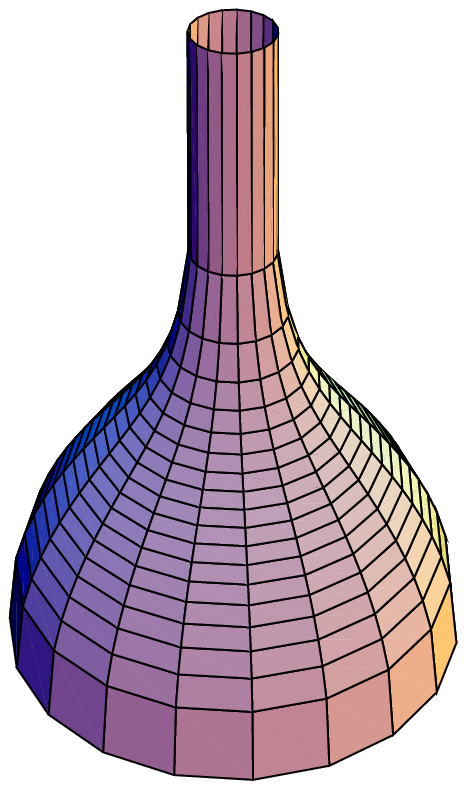}
   \caption{the bounce solution for different background $f=0.6$. The
 radius of the spout of the funnel is the same as that of Fig.\
 \ref{figf-1}, though the radius of the expanded large sphere is
 different due to the difference of the background RR field strength.}
   \label{figf-2}
\end{center}
\end{minipage}
\end{center}
\end{figure}

Another integration constant ${c_4}$ can be determined from the initial
boundary condition. Observe that for a given $f$ and a given 
$N_{\rm D0}$, 
$\p_r T$ of the solution (\ref{soldeca}) 
diverges at two distinct radii $r_1$ and $r_2$ ($r_1 < r_2$). 
This can be seen explicitly in
numerical evaluation. Therefore, the solution (\ref{soldeca}) is
actually a funnel-type which connects two $S^2$'s smoothly along the
Euclidean time direction. One can see this more concretely 
by integrating the expression (\ref{soldeca}) by $r$. The result is
shown in Fig.\ \ref{figf-1} and Fig.\ \ref{figf-2}.
The smaller radius $r_1$ must be the radius of the dielectric brane
$r_{\rm M}$ which is given in (\ref{myr}). 
Hence the equation $\p_rT=\infty$ (which is equivalent to the divergence
of the denominator of (\ref{soldeca})) should be satisfied with 
$r=r_1 \equiv r_{\rm M}$, as
\begin{eqnarray}
 ({c_4})^2 + {c_4} \left(\frac{-4fr_M^3}{3}\right)
+ \frac{4f^2 r_M^6}{9} -r_M^4 - {c_5}^2 =0.
\label{cpeq}
\end{eqnarray}

\noindent
Interestingly, 
two solutions for ${c_4}$ of this equation 
are positive and negative, respectively. 
This can be seen from the fact that if we
substitute ${c_4}=0$ on the left hand side of (\ref{cpeq}) 
it is always negative.
For the negative ${c_4}$,  $T(r)$ is
monotonically decreasing as seen from (\ref{soldeca}), and 
we obtain a funnel-type brane configuration for the bounce, as shown in
Fig.\ \ref{figf-1} and \ref{figf-2}. 
(Note that our funnel is upside-down, since we have
assumed $f>0$ for simplicity. In fact, 
this choice of the orientation of the funnel helps one
to compute the Chern-Simons term of the Euclidean action.) On the other
hand, if ${c_4}$ is positive, $\p_r T$ is positive for small $r$ and
becomes 
negative for large $r$. In this case there 
is a turning point for a certain $r$,
and this results in an interesting brane configuration. In the next
subsection, this solution is identified with the bounce solution for the
nucleation of a dielectric brane from nothing.

In this subsection we use the negative solution for ${c_4}$. Note that,
using this ${c_4}$ we can solve the equation $\p_r T=\infty$ and get
another solution of $r_2$ which is different from $r_{\rm M}$.
This radius is that of the nucleated spherical D2-brane, and at that
radius the expansion starts.
In addition, this nucleated large spherical D2-brane has 
the same D0-brane 
charges as that of the original dielectric brane with $r=r_M$, since
the computation (\ref{evaf}) can be performed independent of its
radius. Thus, we can say that throughout this nucleation and decay the
charge of the D0-branes are conserved. The expansion is dictated by a
hyperboloid worldvolume as in Fig.\ \ref{fige}.
The original D0-branes are brought away to the spatial infinity. 

Fig.\ \ref{figf-1} indicates that the solution of this funnel type is
actually a combination of the curled-up brane and a spike. For a large
$r$, the brane configuration approaches to a hemisphere, while for a
small $r$ the brane configuration can be approximated by a spike, or
rather to say, a throat (volcano) 
solution of a D3-brane found in \cite{Callan}.  
Here we have added newly the magnetic field, thus this throat part is a 
generalization of the solution constructed in \cite{Savvidy} which
is electrically charged.

The equation (\ref{soldeca}) can be integrated as
\begin{eqnarray}
 T =\int_{r_D}^r dr \frac{\displaystyle-\frac{2fr}{3}
+ \frac{{c_4}}{r^2}}
{\sqrt{1-\displaystyle\frac{4f^2 r^2}{9}+\frac{4f {c_4}}{3r}
+ \frac{1}{r^4} \left({c_5}^2 - {c_4}^2\right)}},
\end{eqnarray}
where we have used the boundary condition $T(r_2)=0$ to fix  
zero of time. The Euclidean action is then evaluated as
\begin{eqnarray}
 S_{\rm E} = T_{\rm D2} 
{\displaystyle\int_{r_M}^{r_2}4 \pi r^2 dr
\left(
\frac{1+ \displaystyle\frac{{c_5}^2}{r^4}}
{\sqrt{1-\displaystyle\frac{4f^2 r^2}{9}+\frac{4f {c_4}}{3r}
+ \frac{1}{r^4} \left({c_5}^2 - {c_4}^2\right)}}
-2fT
\right)
}.
\label{finalS}
\end{eqnarray}
Note that in order to obtain the decay width 
we have to subtract the original action of the dielectric
brane. Finally the decay width is given by
\begin{eqnarray}
\Gamma \propto \exp\left(
-\left[
 2 S_{\rm E} - 2 T(r_{\rm M}) E_{\rm dielectric} \right] 
\right)
\label{gammad}
\end{eqnarray}
where
\begin{eqnarray}
E_{\rm dielectric} =
4\pi T_{\rm D2} \sqrt{r_M^4 + 
\pi^2 {\alpha'}^2N_{\rm D0}^2}.
\end{eqnarray}
Here the factor 2 in (\ref{gammad}) 
is coming from the enveloping as before. 
$T(r_{\rm M})$ in front of $E_{\rm {dielectric}}$ is the ``height'' of
the configuration in the Euclidean spacetime. Note that there is no
potential term for $E_{\rm {dielectric}}$. This is because the potential
energy from the background for the dielectric brane was already
subtracted in the expression (\ref{finalS}) since the integration region
in (\ref{finalS}) starts from $r_{\rm M}$, not from $0$. 

To give an analytic expression for the decay width is difficult,
thus here we shall give some numerical results. In the unit
$2\pi{\alpha'}=1$,  
the decay width depending on the background RR field strength is give
as follows. For $N_{\rm D0}=1$, the magnitude of the exponent in the
decay width formula (\ref{coled}) is given as

\begin{figure}[tdp]
\begin{center}
\begin{minipage}{120mm}
 \begin{center}
   \leavevmode
   \epsfxsize=70mm
\put(-60,32){$N_{\rm D0}=0 \longrightarrow$}
\put(-60,46){$N_{\rm D0}=1 \longrightarrow$}
\put(-60,105){$N_{\rm D0}=5 \longrightarrow$}
\put(0,125){$E/(4\pi T_{\rm D2})$}
\put(210,32){$r$}
   \epsfbox{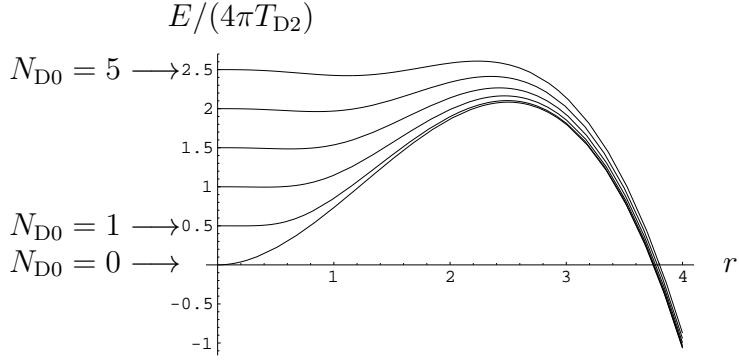}
   \caption{The energy potential (\ref{myene}) 
    for various D0-brane charges ($N_{\rm D0}= 0,1,\cdots,5$). The
  background field strength is fixed ($f=0.4$ in the unit $2 \pi
  {\alpha'}=1$). } 
   \label{figg}
  \end{center}
\end{minipage}
\end{center}
\end{figure}

\vspace{-3mm}
\begin{center}
\begin{tabular}{|c||c|c|c|c|c|c|c|c|c|c|c|c|c|c|}
\hline
$f$ &0.99 &0.9 &0.8 &0.6 &0.4 &0.2 &0.1 &0.01 \\
\hline
$-{\rm exponent}/T_{\rm D2}$ 
&0.129 &3.074 &9.650 &45.89 & 
213.1 &1988 &16466 &1.6653$\times10^7$ \\\hline
\end{tabular}
\end{center}
Note that $f=1$ in the unit $2\pi{\alpha'}=1$ is the critical value for
the existence of the dielectric brane with $N_{\rm D0}=1$, see
(\ref{econ}). For comparison, we give the results for $N_{\rm D0}=0$ as  

\vspace{-3mm}
\begin{center}
\begin{tabular}{|c||c|c|c|c|c|c|c|c|c|c|c|c|c|c|}
\hline
$f$ &0.99 &0.9 &0.8 &0.6 &0.4 &0.2 &0.1 &0.01 \\\hline
$-{\rm exponent}/T_{\rm D2}$ &17.16 &22.85 &32.53 &77.11 &
260.2 &2081 &16655 &1.6655$\times10^7$ \\\hline
\end{tabular}
\end{center}

The results lead us to the following observation : Existence of
D0-branes makes the decay easier. This is very natural, since the
potential barrier in the tunneling is lower for a larger
number of D0-branes. 
The form of the potential for the dielectric branes with various
D0-brane charges is shown in Fig.\ \ref{figg}.

\subsection{Nucleation of dielectric brane from nothing}

\label{nuno}

In this subsection let us briefly discuss what is the positive ${c_4}$
solution of (\ref{cpeq}). As mentioned in the previous
subsection, for the positive ${c_4}$, the derivative of the field $T$
has two regions: A certain value $r_{\rm c}$ exists with which in the
region $r_M<r<r_{\rm c}$, $\p_r T$ is positive, while in the region
$r_{\rm c}<r<r'_2$,  $\p_r T$ becomes negative. This $r'_2$ is the
solution of the equation $\p_r T = \infty$ with given positive ${c_4}$.

\begin{figure}[ht]
\begin{center}
\begin{minipage}{100mm}
 \begin{center}
   \leavevmode
   \epsfxsize=70mm
\put(190,100){$\uparrow -T$}
   \epsfbox{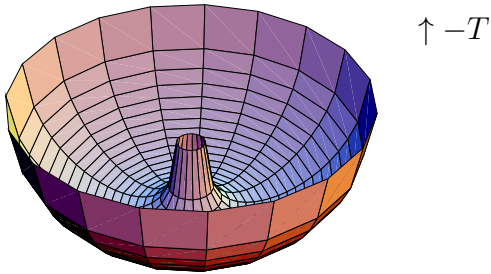}
   \caption{A bounce solution for a positive ${c_4}$ representing a
    nucleation of a dielectric brane and a large spherical brane. In
    this figure we plot $-T(r)$ in order to show inside the doughnut. We 
    have chosen $f=0.5$ and $N_{\rm D0}=1$.}
   \label{fig9}
 \end{center}
\end{minipage}
\end{center}
\end{figure}

The shape of the Euclidean brane configuration is given in Fig.\
\ref{fig9}. The shape is interesting by itself, a half slice of a
doughnut surface. 
If we envelop it with the other half, then we obtain a
complete doughnut surface. 
Note that the heights of the two
$S^p$'s at $r=r_M$ and $r=r'_2$ may not be the same. A numerical
computation generically indicates the heights are in the relation 
$T(r_M)-T(r'_2)>0$. This means, in Fig.\ \ref{fig9} 
where we plot the half doughnut brane upside-down, the central peak is
lower than the surrounding edge of a hemisphere.
Thus to obtain a consistent doughnut after the
enveloping, that is, to adjust two heights, one has to add a
time-independent dielectric brane of the Euclidean time length 
$T(r'_2)-T(r_M)$ on top of the central peak whose shape is a 
small $S^2$ at $r=r_M$.  
This additional configuration would be shown as a cylinder in Fig.\ 
\ref{fig9}. On the other hand, the brane configuration obtained here may 
be understood again as a 
combination of the curled-up brane and a spike (throat). The orientation
of the curled-up brane is opposite to the one given in the previous
subsection. 

The interpretation of this configuration is
much more interesting. This bounce solution represents nucleation of
a pair of spherical branes, as seen from its Euclidean time slice.
Two spherical branes necleated 
have the radii $r=r_M$ and $r=r'_2$, and the
former is a dielectric brane with the D0-brane charge $-N_{{\rm D}0}$
while the latter is a spherical brane which is about to decay, with
an opposite D0-brane charge $N_{{\rm D}0}$. The reason why the former
brane has the negative D0-brane charges is that the orientation of
this brane surrounding the origin $r=0$ is opposite to the latter.
Since the computation (\ref{evaf}) does not depend on the radius to be
used, the total D0-brane charge is conserved. 

The outer spherical D2-brane would decay promptly with expansion, 
while the inner
dielectric brane is meta stable and thus remain for a while. Since we
have not considered the back reaction of the whole system, this
dielectric brane is meta-stable, however if the whole system were in
a 3 dimensional target space then the expansion of the outer brane may
sweep out the background RR field and then the inner brane
may collapse to form point-like $N_{\rm D0}$ D0-branes.

The computation of the decay width is straightforward, and we shall
not give it here. Naively this decay rate must be slightly larger than
the ordinary decay with no D0-brane charge given in Sec.\ \ref{nuex}. 
In Sec.\ \ref{anleave}, a 
similar situation will be found, where a pair of
a brane and an antibrane decays with leaving a lower-dimensional
branes.


\section{Decay of brane-antibrane}

\subsection{Introductory remarks}

The philosophy that all the brane physics in string theory can be
deduced from brane-antibrane pairs via tachyon condensation 
has been supported successfully in K-theory and Sen's conjecture
\cite{Sentac1}. 
On the basis of this, the investigation of the physics of the 
brane-antibrane system is now promoted to the dynamical level; that is,
understanding of the dynamics of the tachyon decay.
Since the annihilation of the brane-antibrane pairs is 
utilized for constructing cosmological inflation 
models of the braneworld \cite{inflation}, 
it is of importance to study how branes annihilate with antibranes 
in various situations in string/M theory.

\begin{figure}[tdp]
\begin{center}
\begin{minipage}{70mm}
\begin{center}
   \leavevmode
   \epsfxsize=50mm
\put(-5,30){D5} 
\put(85,0){D5} 
\put(125,5){D5} 
\put(40,160){D5} 
  \epsfbox{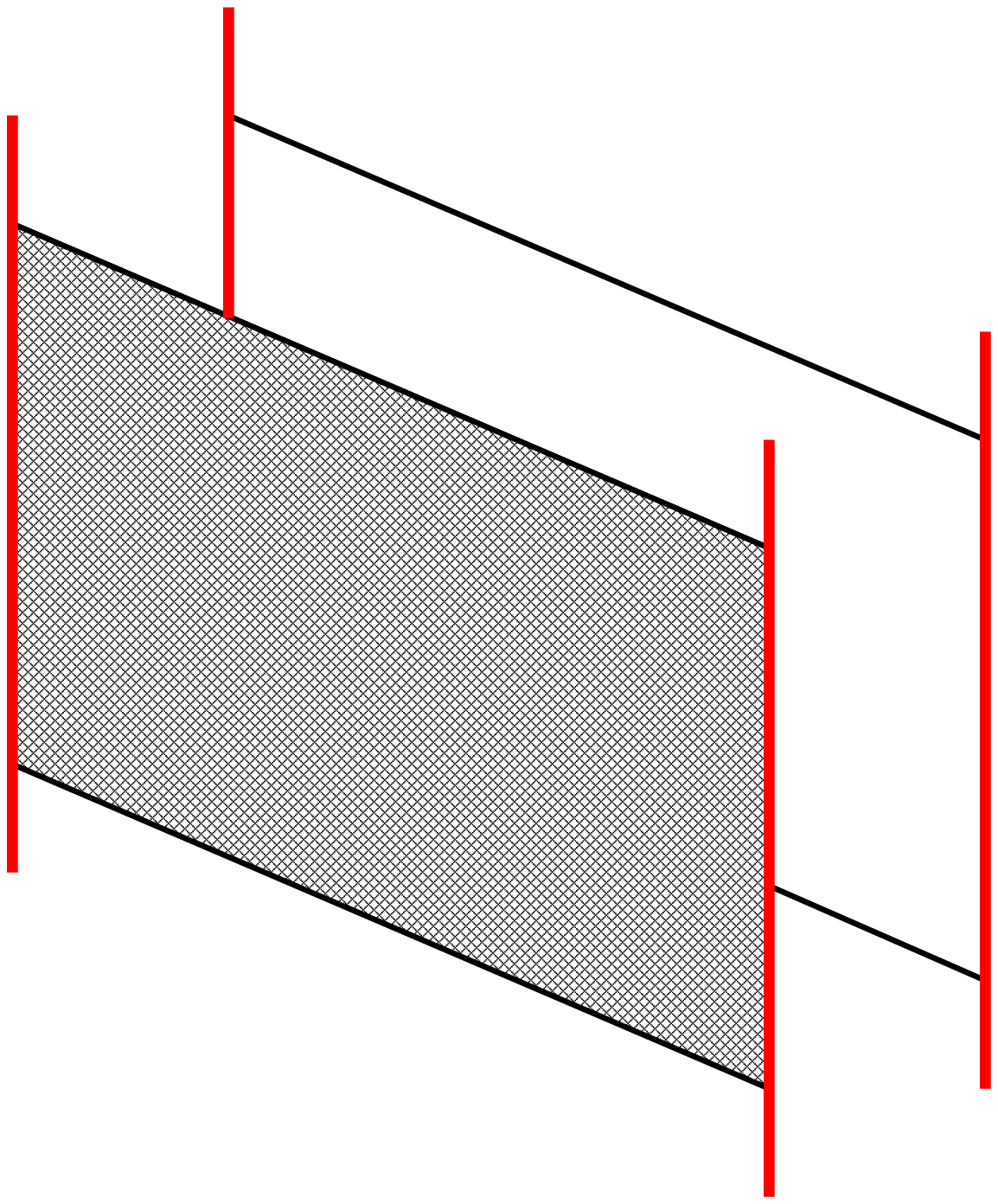}
   \caption{An example configuration in which the asymptotic distance
     between the brane and the antibrane is fixed. Four vertical lines
     are D5-branes on which the D3-brane and the anti-D3-brane have
     ends. } 
   \label{fig1-1}
\end{center}
\end{minipage}
\hspace{5mm}
\begin{minipage}{70mm}
\begin{center}
   \leavevmode
   \epsfxsize=50mm
   \epsfbox{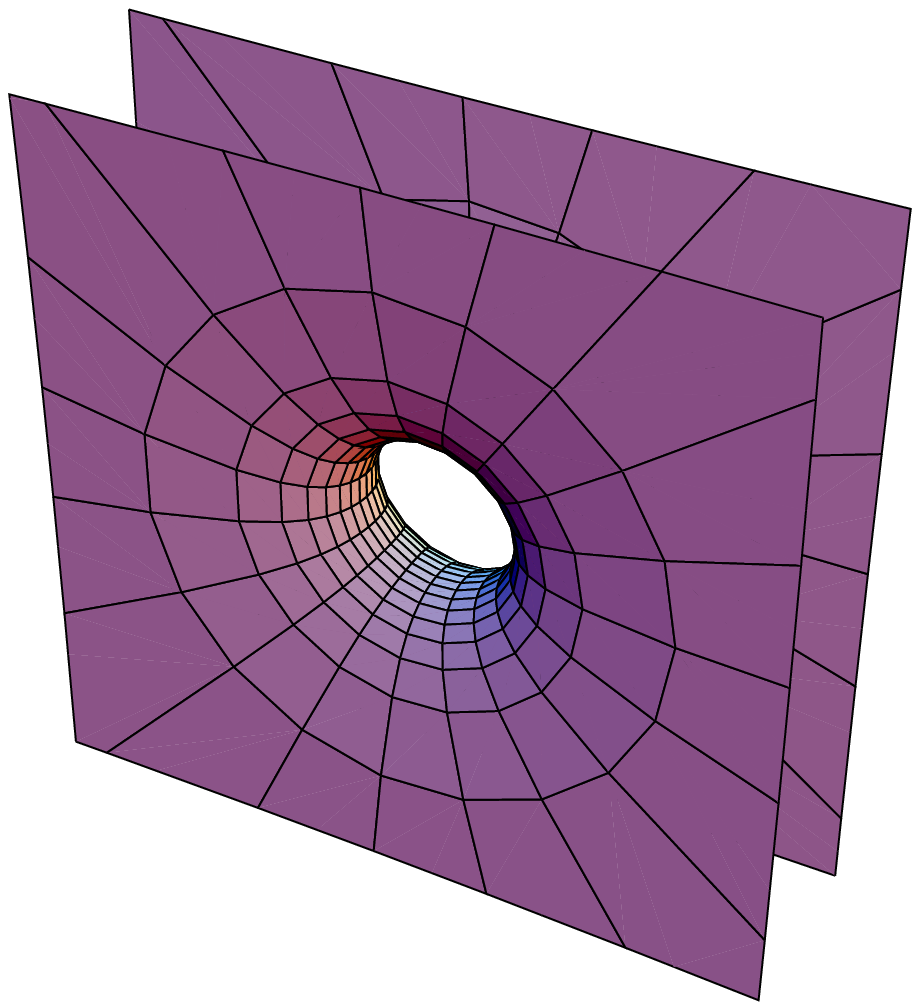}
   \caption{The ``throat'' which connects the brane and the antibrane
     is formed. The decay proceeds with the expansion of the radius of
     this throat. } 
   \label{fig1-2}
\end{center}
\end{minipage}
\end{center}
\end{figure}

Most of the study on the annihilation for the cosmological models 
has been made by use of the gravitational force between the branes.
On the other hand, it has been reported that brane-antibrane pair
can decay in another way: through the tunnel effect by creation 
of a throat between the brane and the antibrane \cite{Callan, 
Savvidy}. If the branes are free to move in the space, 
the time scale of the gravitational approach for the decay is much
smaller than that of the decay via the tunnel effect, as noted in
\cite{Callan}.  
However, we place emphasis on the point that the final stage of the
brane-antibrane decay must be the tachyon condensation since the
pair annihilation and the vanishing of the tension are dictated
by the tachyon condensation. Furthermore, 
there are certain situations where the latter time scale is 
dominant. Consider a brane configuration in which the asymptotic
locations of the brane and the antibrane are fixed in such a way that
the distance between the two cannot change (Fig.\ \ref{fig1-1}). 
Then these branes cannot approach each other by the gravitational
force, hence the dominant time scale for the decay 
may be by the nucleation of the
throat and the expansion of the throat to make the brane-antibrane
vanish (see Fig.\ \ref{fig1-2}).

The decay of unstable systems through the tunnel effect 
may be characterized by the height of
the potential barrier. In the decay of the brane-antibrane, the
top of the potential must be a non-trivial brane configuration.
(For the
decay of the dielectric brane, this is the spherical brane
configuration of the point A' in Fig.\ \ref{figa}.) 
This was explicitly
constructed in \cite{Callan,Savvidy}, and called sphaleron
solutions.\footnote{This is used in a broad sense, and  
slightly different from the original definition of the sphalerons
\cite{spha}.}
These sphalerons were 
solutions in the Born-Infeld-Higgs system on a {\it single} D-brane. 
The scalar field configuration of their solution shows 
a shape of a volcano, a ``half'' of the throat attached to the body of
the flat D-brane.

However, as mentioned, 
the main contribution to the brane-antibrane decay 
must be the condensation of the
tachyon which comes from an excitation of a string connecting the
brane and the antibrane.  
In this section, we reconsider this formation of the throat from the
viewpoint of the tachyon condensation. The formation of the
throat is the result of the quantum 
tunnel effect. We investigate this tunnel
effect and compute its probability, by using the effective field
theory of the tachyon fields living on the brane-antibrane. 
We show the existence of a bounce solution in the tachyon field
theory, and the brane configuration formed by this bounce solution
as its final state is found to be a throat-like. 
Inside the throat, the
tachyon sits at the true vacuum known as the
closed string vacuum, thus there
is no brane there. The throat itself is a spherical bubble made of a
domain wall. 
Outside the bubble the tachyon expectation value is zero, thus the
brane and the antibrane has not vanished yet there. 
After the nucleation of the throat, the tachyon configuration ``rolls
down'' the potential hill. 
\footnote{
  This is inhomogeneous roll down which is different from the recent 
  computation using a homogeneous ansatz \cite{Senroll}.} 

The analysis in this section proceeds in the following way.
First, we derive the action of the relevant tachyon system
of the separated brane-antibrane. We find that the parallel
brane-antibrane corresponds to a false vacuum (a local minimum of the  
tachyon potential) of the tachyon field theory. We apply Coleman's
method for describing the decay of the false vacuum via the tunnel
effect, and construct a relevant bounce solution for the Euclidean 
tachyon field theory. We show that the Euclidean action evaluated with
this bounce solution is reproducing a naive energy of a throat of the
Euclidean brane computed by its tension times the area (volume). 
The radius $R_0$ 
of the throat of the bounce solution is at the order of the
brane separation, $R_0 \sim l$. 
Interestingly, the energy of the throat is very similar to what was
found in Sec.\ \ref{nuex}. 
Instead of the background RR field strength $f$ in Sec.\ \ref{nuex},
here we have the brane separation parameter $l$. The correspondence is
almost $l \sim f^{-1}$. We may use this similarity to argue how the
decay proceeds after the nucleation of the throat.


\subsection{Naive prospect}

Before going to the detailed study of the decay width using the
tachyon condensation, here we shall give a naive estimation of the
decay width. This naive prospect is based only on the tensions and
shapes of the branes. Later we will see that this prospect coincides
with the estimation by the tachyon condensation. 

\begin{figure}[tbp]
\begin{center}
\begin{minipage}{120mm}
\begin{center}
   \leavevmode
   \epsfxsize=50mm
   \epsfbox{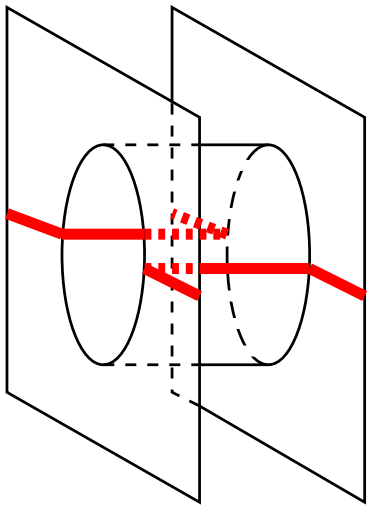}
   \caption{A naive expectation of the bounce solution. Here we draw a
    figure for $p=1$. The throat is a cylinder, and it makes a hole in
    each Euclidean D1-brane worldvolume. The thick lines denote a
    Euclidean time slice $\tau=0$, which represents a nucleated throat
    between the two D1-branes. }  
   \label{fignaive}
\end{center}
\end{minipage}
\end{center}
\end{figure}

First, the tunnel effect is caused by the bounce solution which has
the highest symmetry in the Euclideanized theory and satisfies
appropriate boundary conditions. In our case, starting from the
parallel D$p$-brane and anti-D$p$-brane, we expect that the
resultant brane configuration after the tunneling will be a throat
connecting two flat branes. This is shown in Fig.\ \ref{fignaive}. 
Therefore, the bounce solution should have the Euclidean
(p+1)-dimensional worldvolume, and thus this bounce configuration is
effectively a static configuration of a 
D$(p+1)$-brane. As shown in Fig.\ \ref{fignaive}, 
the nucleated throat is found in the $\tau=0$ (Euclidean) time slice. 
According to Coleman's method which will be explained and generalized
in detail in the next subsection, the decay width by this bounce
solution is given by (\ref{coled}). 
In our case, as we stated before, the Euclidean action which is to be
substituted is just the Euclidean worldvolume area times the brane
tension. Thus, for the brane configuration given in Fig.\
\ref{fignaive},  the action is given as
\begin{eqnarray}
  S_{\rm E} = 
\left[
  A(S^p) l R_0^p - 2  V(S^p)R_0^{p+1}
\right]T_{{\rm D}p}
=\left[
 A(S^p) R_0^p  - 2l^{-1}  V(S^p)R_0^{p+1}
\right]T_{{\rm D}p} l.
\label{naiveac}
\end{eqnarray}
Note that the first term is the action of the throat whose section is
$S^p$,  and the second term is for two holes in the parallel
brane-antibrane. 
The bounce solution must be the extremum of the action, thus we
can determine the radius $R_0$ by $\delta S_{\rm E}/ \delta R_0=0$,
which gives 
\begin{eqnarray}
  R_0 = \frac{p}{2}l.
\label{rnaive}
\end{eqnarray}
Interestingly, the expression (\ref{naiveac}) 
is very similar to what we encountered
in the evaluation of the action for the nucleation of the spherical
branes in the constant RR field strength (\ref{actionde}). 
This analogy will be checked further  by confirming 
the expression (\ref{naiveac}) in the tachyon condensation in the
following subsections. The intuition of the analogy will help 
us to find the decay width of the brane-antibrane while leaving a
lower dimensional branes in Sec.\ \ref{anleave} and to understand how
the decay proceeds after the nucleation of the throat.

\subsection{Description of brane-antibrane by tachyon field }

Let us state our situation more precisely : 
a D$p$-brane and an anti-D$p$-brane are
sitting with a distance $l$, in parallel. 
If the distance $l$ is large enough, then the tachyon mode between the
two should go away, since the complex tachyon is coming from the string
suspended between the two branes and thus that string acquire a mass
lift when two branes are distant. In fact, the mass of the 
lowest excitation mode of the string connecting these branes is given by 
\begin{eqnarray}
 m^2 = \left(
\frac{l}{2\pi{\alpha'}}
\right)^2
-\frac{1}{2{\alpha'}}.
\label{mass2}
\end{eqnarray}
Therefore evidently this tachyonic mode becomes massive when the
separation $l$ becomes larger than $\sqrt{2\pi^2{\alpha'}}$. 

However, intrinsically the system we are considering must be unstable,
since if these two branes annihilate then the total energy of the system 
will decrease. Therefore, the stability of the separated brane-antibrane
system considered above must be perturbative, and we expect that the
true vacuum of the system will be obtained
nonperturbatively.\footnote{Note that brane-antibrane system with
  an appropriate electro-magnetic field strength can be supersymmetric
  and hence stable. See \cite{bpsba}.}
Although perturbative tachyon potential has been obtained 
from the string scattering amplitude in this background brane-antibrane
\cite{Pesando},  
we need nonperturbative information on the tachyon potential. 
Fortunately, exact tachyon potentials have been computed in the boundary 
string field theory (BSFT) \cite{BSFT}, 
and a part of Sen's conjectures have been verified \cite{verify}. The
tachyon effective field theories obtained as a derivative truncation of
BSFT actions are found to reproduce a partial spectra of D-branes from 
tachyon defects \cite{MZ}. We shall use this tachyon effective action
for our investigation in the following.

In order to obtain the brane-antibrane configuration to be dealt with,
first consider the D$(p\!+\!1)$-brane - anti-D$(p\!+\!1)$-brane 
on top of each other. The action is given by \cite{Terashima} as
\begin{eqnarray}
 S = T_{{\rm D}(p+1)}\int\! d^{p+2}x\;  e^{-|T|^2}
\left[1+2{\alpha'}|D_\mu T|^2  
+\frac{(\pi{\alpha'})^2}4 (F_{\mu\nu}^{(-)})^2
\right],
\label{action1}
\end{eqnarray}
where we employed the two-derivative truncation of the BSFT.
The complex tachyon field $T$ is charged under the gauge group $U(1)$
associated with the gauge field 
\begin{eqnarray}
 A^{(-)}_\mu \equiv A^{{\rm D}(p+1)}_\mu  - 
A^{{\rm antiD}(p+1)}_\mu,
\end{eqnarray}
and we ignored another combination of the gauge fields since it is
irrelevant for our following discussion.
Now, turn on the constant Wilson line $A_{p+1} = A$ 
for this gauge field and take a T-duality along $x_{p+1}$ axis. 
Then, the gauge field
is mapped to the Higgs field which measures the distance between the
two, and the distance is given by the absolute value of the Wilson
line.
If we neglect all the fields other than the tachyon field, then we
obtain an action
\begin{eqnarray}
 S = 2T_{{\rm D}p}\int\! d^{p+1}x\;  
e^{-|T|^2}
\left[
1+2{\alpha'}|\p_\mu T|^2  +2 {\alpha'}A^2 |T|^2
\right].
\label{action2}
\end{eqnarray}

\begin{figure}[tdp]
\begin{center}
%
   \leavevmode
   \epsfxsize=70mm
\put(20,120){$V(T)$}
\put(180,20){$T$}
   \epsfbox{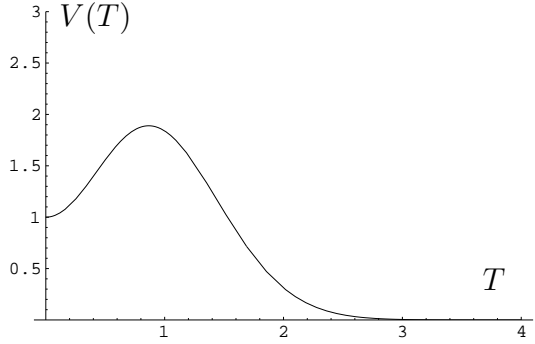}
   \caption{Potential (\ref{potential}) with $A=2$ and $2{\alpha'}=1$.}
   \label{fig1}
\end{center}
\end{figure}

It is very easy to see
that this action has the desired properties discussed
above. The potential term of this action has an expansion
\begin{eqnarray}
 V(T) = 
e^{-|T|^2}
\left[
1+2{\alpha'}A^2 |T|^2
\right]
=1+ \left(2{\alpha'}A^2-1 \right)|T|^2 + {\cal O}(T^4),
\label{potential}
\end{eqnarray}
Thus the mass for this tachyon field is given by
\begin{eqnarray}
 m^2 = A^2-\frac{1}{2{\alpha'}}.
\end{eqnarray}
One can see the precise correspondence between (\ref{mass2}) and this
equation, and we have 
\begin{eqnarray}
A=\frac{l}{2\pi{\alpha'}}.
\end{eqnarray}
Furthermore, the above
potential tells us that, if $A$ is large enough then the system is
perturbatively stable at $T=0$, but the true vacuum of the system is
lying at $T=\infty$ which is so-called closed string vacuum, i.e. where
two D$(p\!+\!1)$-branes 
do not exist. The profile of the potential $V(T)$ is
shown in Fig.\ \ref{fig1}. 

{}From this respect, the annihilation of two distant branes is just the
transition from the false vacuum $T=0$ to the true vacuum $T=\infty$,
through the potential barrier. Let us consider the properties of this
potential barrier. The height of the potential is at $T_{\rm top}=
\sqrt{1-1/(2{\alpha'}A^2)}$,
and if we assume that $l \gg l_s \equiv \sqrt{{\alpha'}}$ 
where $l_s$ is a string length, 
this is approximated by 
$T_{\rm top}\sim 1$. The potential hight at this point is 
$V_{\rm top} \sim 2{\alpha'} A^2/e$. 
Thus if the distance between the two branes
become large, the potential barrier grows (as $\sim l^2$). 
This is very natural. For a possible BSFT correction to this tachyon
potential, see Appendix A.


\subsection{Decay width of brane-antibrane}


\subsubsection{Coleman's computation in arbitrary dimension}

The prescription on how to compute the decay width of the false vacuum
was given by S.\ Coleman \cite{coleman1}. Here we summarize his results
\cite{coleman1, Coleman}
and generalize it to the case of field theories with arbitrary
spacetime dimensions. The system considered in \cite{coleman1} 
is a scalar field theory with an action
\begin{eqnarray}
 S = \int d^{p+1}x 
\left[- \Half \left(\p_\mu\Phi\right)^2+U(\Phi)\right].
\label{usual}
\end{eqnarray}
Here we have generalized it to the system of $p+1$ dimensional
spacetime, and $U(\Phi)$ is assumed to have a false vacuum. One may
imagine the potential of Fig.\ \ref{fig1} as $U(\Phi)$. 
Suppose that one starts with a homogeneous false vacuum. Then the
tunneling toward the true vacuum is described by a bounce
configuration which is a solution of an Euclidean equation of motion.
The principle to have that bounce solution is the highest symmetry, 
as mentioned in the previous section, that preserves all of the
boundary conditions. In our case, the highest symmetry is the
rotational invariance $SO(p+1)$ in the Euclidean space. Thus the
bounce solution is a true vacuum bubble. The surface of the bubble
consists is a domain wall, and the inside of the bubble is true vacuum
while the outside sits still at the false vacuum.
The decay width is given in the formula (\ref{coled}).
For a high potential barrier, we may adopt the thin wall
approximation, then the result is
\begin{eqnarray}
 S_{\rm E} &=& \int_0^\infty A(S^p) \rho^p 
\left[\Half \left(\frac{d\Phi}{d \rho}\right)^2+U(\Phi)\right] d \rho
\nn\\
&=&  A(S^p) R^p S_1-V(S^p) R^{p+1} \epsilon
\end{eqnarray}
where $R$ is the radius of the bubble estimated later. In this
approximation, the ``wall energy'' $S_1$ is approximated by that of
the  one dimensional system, 
\begin{eqnarray}
 S_1 \equiv \int_{\Phi_0}^{\Phi_1}d\Phi \sqrt{2U(\Phi)}.
\end{eqnarray}
Here $\Phi_0$ is the value of the field 
giving the false vacuum of the potential $U(\Phi)$, while $\Phi_1$ is
that for the true vacuum. 
The parameter $\epsilon$ is the difference of the energies of the two
local minima,
\begin{eqnarray}
 \epsilon \equiv U(\Phi_0) - U(\Phi_1).
\end{eqnarray}
The thin wall approximation is applicable if \cite{coleman1}
\begin{eqnarray}
 \frac{S_1 \mu}{\epsilon} \gg 1,
\label{valid}
\end{eqnarray}
where $\mu$ is the mass scale of the system. This $\mu$ is given by the
coefficients of the mass term evaluated at the top of the potential.

Since the energy of the bubble is written by its radius, we can
determine the radius of the bubble as a solution of the equation
\begin{eqnarray}
 \frac{\delta S_{\rm E} }{\delta R} =0.
\end{eqnarray}
For arbitrary dimensions, we obtain
\begin{eqnarray}
 R = \frac{p}{p+1} \frac{A(S^p)}{V(S^p)}\frac{S_1}{\epsilon}.
\label{radiko}
\end{eqnarray}
Then, using this radius, the Euclidean action is finally evaluated as
\begin{eqnarray}
 S_{\rm E} = \frac{p^p}{(p+1)^{p+1}}
\frac{(A(S^p))^{p+1}}{(V(S^p)^p)}\frac{(S_1)^{p+1}}{\epsilon^p}
= \frac{p^p}{p+1}
\frac{2 \pi^{(p+1)/2}}{\Gamma\left(\frac{p+1}{2}\right)}
\frac{(S_1)^{p+1}}{\epsilon^p}.
\end{eqnarray}
In the next subsection, we apply this result for the decay of the
brane-antibrane.


\subsubsection{Application to ours}

To apply the generalized Coleman's computation to our situation, there
exists two obstacles coming from the difference between our action
(\ref{action2}) and a usual action (\ref{usual}). First, our system
has a complex tachyon field while (\ref{usual}) is a real scalar
field. Second, the tachyon action (\ref{action2}) has an unusual
kinetic term. 

\begin{figure}[tdp]
\begin{center}
\begin{minipage}{120mm}
\begin{center}
   \leavevmode
   \epsfxsize=70mm
   \epsfbox{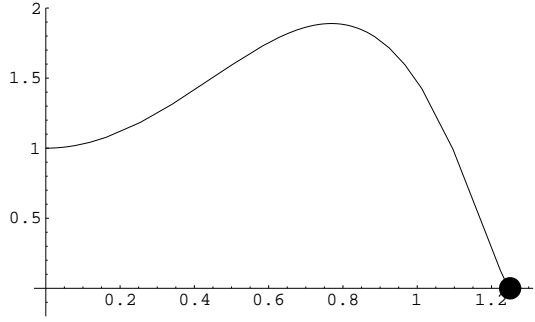}
   \caption{The potential $U(\Phi)/2T_{{\rm D}p}$ with $A=2$,
     $2{\alpha'}=1, 8{\alpha'} T_{{\rm D}p}=1$ for simplicity. 
   The true vacuum (depicted as a blob) is located at the finite
   distance $\Phi_1$.} 
   \label{fig4}
\end{center}
\end{minipage}
\end{center}
\end{figure}

For the first difference, here we simply assume that the relevant
tachyon bounce solution is real. The vanishing of the imaginary part
of the tachyon is consistent with the tachyon equations of motion, 
and furthermore 
this is a very natural ansatz since we are considering
no lower dimensional D-brane left after the tachyon condensation. 
(The lower dimensional D-branes which are stable 
may appear after the tachyon condensation as topological defects
\cite{Sentac1}, and this needs complex tachyon configurations.)
For the real tachyon field,
\begin{eqnarray}
S = 2 T_{{\rm D}p}    \int d^{p+1}x e^{-T^2} 
\left[1+2{\alpha'} (\p_\mu T)^2 + 2{\alpha'} A^2 T^2\right].
\end{eqnarray}
Then, with the real tachyon, we can redefine the tachyon field so that
the action have a canonically normalized kinetic term: 
\begin{eqnarray}
 \Phi \equiv \sqrt{8 {\alpha'} T_{{\rm D}p}} \int_0^T ds \; e^{-s^2/2}.
\label{redef}
\end{eqnarray}
Note that, although the original true vacuum of the tachyon is at the
infinity $|T|=\infty$,  this redefinition makes it to be placed at a
finite value of the new field $\Phi$. In fact, two local minima are
\begin{eqnarray}
 \Phi_0 = 0, \quad 
\Phi_1 = \sqrt{4 \pi {\alpha'} T_{{\rm D}p}}.
\end{eqnarray}
However, the field redefinition does not change the energy difference
\begin{eqnarray}
\epsilon = 2 T_{{\rm D}p}.
\end{eqnarray}
The form of the potential written by this new field is shown in Fig.\
\ref{fig4}.\footnote{In the figure it looks that around the true
  vacuum the potential is not flat. However, evaluating the derivative
  of the potential, we find that it is actually flat: 
  \begin{eqnarray}
    \frac{d U(\Phi)}{d \Phi}\biggm|_{\Phi=\Phi_1}
\propto \left[T e^{-T^2} \frac{dT}{d\Phi}\right]_{\Phi=\Phi_1}
\propto \left[T e^{-T^2/2}\right]_{T=\infty}=0.
  \end{eqnarray}
}

We need to determine the scale parameter $\mu$ of the theory by 
studying the behaviour of the potential around its maximum. 
The potential written in terms of the tachyon is
\begin{eqnarray}
 U = 2 T_{{\rm D}p}e^{-T^2} (1+2{\alpha'} A^2 T^2).
\end{eqnarray}
For the high potential, we have large ${\alpha'} A^2$, hence
the top of the hill is at $T\sim 1$. There 
\begin{eqnarray}
\frac{\delta^2 U}{\delta T^2}  \sim -2 T_{{\rm D}p}
\frac{8{\alpha'} A^2}{e}. 
\end{eqnarray}
The redefinition (\ref{redef}) at $T=1$ gives
\begin{eqnarray}
 \delta \Phi = \sqrt{8 {\alpha'} T_{{\rm D}p}}\; e^{-1/2} \delta T,
\end{eqnarray}
thus 
\begin{eqnarray} 
U - U_{\rm top} \sim -A^2 (\delta \Phi)^2
\end{eqnarray}
hence the mass scale evaluated at the top of the hill is given by
\begin{eqnarray}
 \mu \sim 2 A.
\end{eqnarray}

Next, we evaluate $S_1$. Writing $S_1$ by integration with $T$, 
\begin{eqnarray}
 S_1 &=& \int _0^{\sqrt{4\pi{\alpha'} T_{{\rm D}p}} }
d\Phi \sqrt{2 U(\Phi)} \nn\\
&=&
\int_0^\infty dT e^{-T^2/2} \sqrt{8 {\alpha'} T_{{\rm D}p}}
\sqrt{2\cdot 2 T_{{\rm D}p} e^{-T^2}(1+2{\alpha'} A^2 T^2)}
\nn\\
&=&
4\sqrt{2{\alpha'}} T_{{\rm D}p}
\int_0^\infty dT e^{-T^2} \sqrt{1+2{\alpha'} A^2 T^2}. 
\label{integ}
\end{eqnarray}
In this expression, it is easy to see the validity of the thin wall
approximation (\ref{valid}), since the left hand side of (\ref{valid})
\begin{eqnarray}
 \frac{S_1 \mu}{\epsilon} = 
4\sqrt{2{\alpha'} A^2}\int_0^\infty
dT e^{-T^2} \sqrt{1+2{\alpha'} A^2 T^2}
\end{eqnarray}
is much larger than $1$  if $2 {\alpha'} A^2$ is large. 
Let us continue the evaluation of $S_1$. We use the steepest-descent
method by defining the integrand as an exponential form,
\begin{eqnarray}
 e^{-T^2} \sqrt{1+2{\alpha'} A^2 T^2}
\equiv e^{f(T)}.
\end{eqnarray}
This $f(T)$ has a maximum at $T \sim \sqrt{1/2}$ where the second
order derivative is  $\delta^2 f / \delta T^2 \sim -4$. Thus
the integration of (\ref{integ}) is evaluated as 
\begin{eqnarray}
  \int_0^\infty dT e^{-T^2} \sqrt{1+2{\alpha'} A^2 T^2}
& \sim & \int_0^\infty dT e^{-1/2}
\sqrt{1+{\alpha'} A^2} e^{-2(\delta T)^2} 
\nn\\
& \sim &
\sqrt{\frac{1+{\alpha'} A^2}{e}} \int_{-\sqrt{1/2}}^{\infty} 
d(\delta T) e^{-2(\delta T)^2}
\nn\\
& \sim & 
c_0 \sqrt{\frac{{\alpha'} A^2}{e}},
\end{eqnarray}
where numerically $c_0 \sim 1.15$.
Therefore, we obtain
\begin{eqnarray}
 S_1 
\; \sim \; 
4 \sqrt{2{\alpha'}}T_{{\rm D}p}  c_0 \sqrt{\frac{{\alpha'} A^2}{e}}
\; \sim \;
0.628 \; T_{{\rm D}p} l.
\end{eqnarray}

Then we obtain the radius (\ref{radiko}) of the nucleated bubble as
\begin{eqnarray}
 R \; \sim\;  \frac{2\sqrt{2} c_0}{2\pi\sqrt{e}}p \; l 
\; \sim \; 0.314 p \; l.
\end{eqnarray}
We see that this radius is given by $\sim l$, 
The naive estimation (\ref{rnaive}) gives a correct dependence on
the brane separation $l$, although we have another dimensionful
parameter $l_s$. 
Finally, the decay width is given by the Euclidean action, 
\begin{eqnarray}
 S_{\rm E} &=& 
\frac{p^p}{p+1}
\frac{1}{\Gamma\!\left(\frac{p+1}{2}\right)}
\frac{1}{2^{p-1} \pi^{(p+1)/2}}
\left(
\frac{2\sqrt{2}\cdot c_0}{\sqrt{e}}
\right)^{p+1}
l^{p+1} T_{{\rm D}p}
\nn\\
&=&
\frac{p^p}{p+1}
\frac{1}{\Gamma\!\left(\frac{p+1}{2}\right)}
\frac{1}{2^{2p-1} \pi^{3(p+1)/2}} 
\left(
\frac{2\sqrt{2}\cdot c_0}{\sqrt{e}}
\right)^{p+1}
\left(\frac{l}{l_s}\right)^{p+1} \frac{1}{g}.
\end{eqnarray}

From the result above, especially the coincidence with the naive
prospect in Sec.\ 3.2, we claim that this bubble of the true vacuum
corresponds to a throat connecting the brane and the antibrane. 
To make sure that this claim is correct, in the next subsection we study
the scalar field configuration with this tachyon bubble, since the shape
of the throat is determined by this scalar field on the
brane-antibrane. 

As we noted, the expression for the Euclidean action (\ref{naiveac}) 
evaluated by the bounce solution is very similar to what we have found
in the nucleation of a spherical D$p$-brane in the constant RR field
strength. Thus from this observation 
it is easy to guess the fate of the nucleated throat in the
brane-antibrane, from Fig.\ \ref{fige}. 
The tachyon at each radius outside of the throat rolls
down the hill in due order, and the throat (which is the tachyon domain
wall) expands. The speed of the
expansion of the throat radius may approach the speed of light, then the
brane-antibrane is swept out to vanish.


\subsection{Inclusion of scalar field} 

Originally the tachyon field is interacting with infinitely many
other fields which are also excitations of the strings. Among them,
massless fields such as scalar fields and vector gauge fields are of
importance, and we had to incorporate these fields to make the 
description precise. However, it is of course difficult to treat all
the modes at the same time, thus we have considered only the tachyon
mode. We have seen that this approximation works well, that is, 
we have found that the decay width of the brane-antibrane is that of
the naive computation by making a throat.

Now let us try to incorporate the scalar field in this system. 
We shall not consider the gauge fields because it is relevant for the
situation where some lower dimensional D-branes are left after the
tachyon condensation. Here we consider only the complete annihilation 
of the brane-antibrane pair. For the case leaving smaller D-branes, see 
Sec.\ \ref{anleave}.

The scalar field is actually what was originally studied in
\cite{Callan, Savvidy}, and in these papers the existence of a kind of
sphaleron (a throat solution) was shown. 
Therefore we expect that incorporation of the 
scalar field $X$ into our system does not change the physical results.
The relevant scalar field here is the combination
$X^{\rm (D4)}-X^{\rm (antiD4)}\equiv X$
which measures the distance between the two branes. This can be easily
seen from the T-duality which we considered in Sec.\ 3.3. 
The value of the field $X$ should be finite. 

The Euclidean action of the scalar field interacting with the tachyon 
is written as
\begin{eqnarray}
  S = 2T_{{\rm D}p} \int \! d^{p+1}x \; 
e^{-|T|^2} 
\left[
\frac{(\pi{\alpha'})^2}2 
(\p_\mu X)^2 + X^2 |T|^2
\right].
\label{intx}
\end{eqnarray}
We employed again an action which is two-derivative truncation of the
BSFT results \cite{Terashima}. 
Since it is difficult to solve the full equations of motion, 
we regard the tachyon solution as an background and consider how the
scalar field behaves on this background.

Therefore, first let us see how the solution of the tachyon equations
of motion behaves in various regions. 
As in the previous sections, we assume that the tachyon field is a
real function of $r$ where $r$ is a radial coordinate of the Euclidean
D$p$-brane worldvolume. We consider the case $p=2$ for
simplicity.\footnote{For $p=2$, the Euclidean worldvolume is three
  dimensional, thus it might be easy to compare it with the static
  ``sphaleron'' solution in \cite{Callan}. } 
Under this assumption, we obtain the equation of motion as
\begin{eqnarray}
 T 
\left(
\frac{1}{2{\alpha'}}-A^2 - (\p_r T)^2 + A^2 T^2
\right)
+ \p_r^2 T + \frac2r \p_r T =0.
\label{eomt}
\end{eqnarray}
The last two terms come from the Laplacian acting on $T$.

\begin{figure}[htdp]
\begin{center}
\begin{minipage}{100mm}
\begin{center}
   \leavevmode
   \epsfxsize=70mm
   \epsfbox{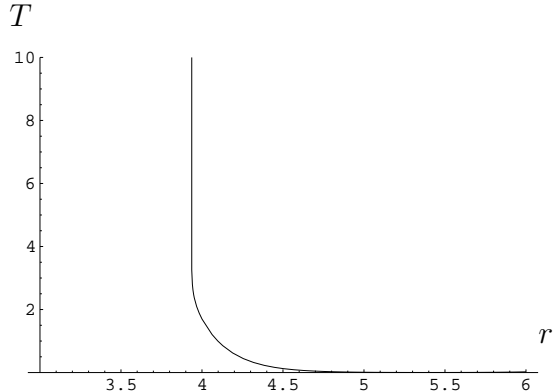}
  \put(-200,140){$T$}
  \put(0,20){$r$}
   \caption{A numerical solution to the equation of motion with 
   $A=5$ in the unit $2{\alpha'}=1$. 
   The boundary condition is chosen as $T(r\!=\!5)=0.01, 
   T'(r\!=\!5)=-\sqrt{A^2-1}\;T(r\!=\!5)$, according to the asymptotic
   solution (\ref{asymp}).}
   \label{fig3}
\end{center}
\end{minipage}
\end{center}
\end{figure}

Let us study the asymptotic behaviour of the solution. 
The physical requirement is that, in the region with large $r$,
$T$ must sit at the false vacuum $T=0$. Neglecting the terms of the
higher order in $T$ and also the last term for the large $r$, we obtain
an appropriate asymptotic solution 
\begin{eqnarray}
T \propto \exp
\left[-r \sqrt{A^2-\frac{1}{2{\alpha'}}}\right].
\label{asymp}
\end{eqnarray}
Using this behaviour 
as an input at large $r$, we can solve the equations of
motion numerically. The result is very interesting, as shown in 
Fig.\ \ref{fig3}. 
One observes that at a certain $R$ the solution $T$
diverges to the infinity. 

Divergence of $T$ is equivalent to the approach to the true vacuum
$\Phi=\Phi_1$. Therefore, the numerical solution in Fig.\ \ref{fig3}
is actually a domain wall solution which interpolates from the false
vacuum to the true vacuum. In the region $r <R $, the tachyon sits at
the true vacuum, and thus the open string physics there is expected to
disappear completely. This means that there is no D2-brane there.
We expect that after the tunneling the D2-branes are connected by
making a throat between them. Inside the throat radius, we have only the 
holes.

Then, let us consider the scalar field behaviour in this tachyon
background of the bounce solution. 
The equations of motion for the scalar field $X$ becomes
\begin{eqnarray}
2e^{-|T|^2}|T|^2  X
-(\pi{\alpha'})^2
\p_\mu \left(
e^{-T^2} \p_\mu X
\right)
=0.
\label{eomp}
\end{eqnarray}
We substitute the bounce solution for $T$. 
Around $r=r_0$ the tachyon $T$ diverges, and generically near the
divergent point $|\p_rT|$ is much larger than $T$ itself. 
Thus we expect that the dominant behaviour of $\Phi$ around $r=r_0$
comes from the second term of the equation above.  
Neglecting the first term, we have
\begin{eqnarray}
\p_r
\left(
e^{-T^2} r^2 \p_r X
\right)
=0.
\end{eqnarray}
Here we have used the assumption that the scalar
field is rotationally symmetric. Then it is easy to find a solution
for the scalar field as
\begin{eqnarray}
 \p_r X = e^{T^2}\frac{c}{r^2}.
\end{eqnarray}
This indicates that the derivative of the scalar field diverges
at $r=R$. This result is very consistent, since from the tachyon 
bounce solution we expected that at $r=R$ the two branes are connected
and making a throat. At that point the gradient of the scalar field must 
be divergent. Thus we have a consistency.

On the other hand, we can solve the equations of motion (\ref{eomp})
asymptotically by substituting the asymptotic tachyon solution
(\ref{asymp}). 
The resultant asymptotic behaviour of $X$ is $\sim 1/r$ which is
the same as what was found in the ``sphaleron'' solution in 
\cite{Callan}. So we have another consistency.
The difference between ours and \cite{Callan} is that the nontrivial
throat part in \cite{Callan} comes from the nonlinearity of the action
in $X$, while the divergence of $\p_rX$ in our case is due to the
coupling with the tachyon.  

Unfortunately, we don't have any indication that the distance between
the two branes which must be given by integrating $\p_r X$ is
actually the parameter $l$. Perhaps in order to see this we have to 
study the full equations of motion of the interacting system including
${\alpha'}$ corrections.\footnote{One may wonder if the inclusion of $X$
crucially affects our result, from the following observation: The
interaction term $X^2 T^2$ in (\ref{intx}) makes the potential barrier
lower if $X$ decreases, and if $X$ become lower than the critical value  
$X_{\rm c}=1$ then the classical instability of tachyon appears in that
region, and thus the analysis only with the tachyon with flxed
brane separation might not be
correct. However, we expect that this is not the case. The reason is as
follows. First, at the point $T$ diverges ($r=r_0$), also $X$ should
diverge, since in the region $r<r_0$ the tachyon sits at the closed
string (true) vacuum and there should be no open string dynamics. Hence
at the boundary $r=r_0$ it is natural to expect that the throat
connecting the brane and the antibrane exists, and to make a smooth
throat $\p_r X$ should diverge at $r=r_0$. This divergence 
means that the region where the tachyon has classically tachyonic mass
squared is very tiny ($r_0 < r < r(X_{\rm c})$). Therefore it may not be 
necessary to take care of the classically tachyonic region near the
throat.   }


\subsection{Annihilation leaving topological defect}
\label{anleave}

Up to here in this section we have assumed a real configuration of the
tachyon field. If we consider a nontrivial winding of the complex
tachyon field onto the infinity of the 2-dimensional subspace of the
worldvolume,  then we may describe a situation where the tachyon
condensation leaves a D$(p\!-\!2)$-brane. 

Let us consider especially the case with $p=2$ for simplicity. 
In this case, a D0-brane may be left after the condensation. 
It is easy to write a tachyon action under the assumption that 
\begin{eqnarray}
  T= t(r) e^{i \theta} 
\end{eqnarray}
where $t(r)$ is a real function of $r\equiv \sqrt{(x^1)^2 + (x^2)^2}$,
and $\tan \theta \equiv x^2/x^1$, which  are the polar coordinates.
This nontrivial winding corresponds to a single D0-brane charge sits
at the origin. The Euclidean action written with this ansatz is
similar to what we have considered for the real tachyon, though the
boundary condition is slightly different: to guarantee the
single-valuedness for the tachyon at the origin $r=0$, we must have
$t(0)=0$. Because of this constraint, there remains a small part of
the false vacuum even inside the throat. This small part is actually
expected to be a D0-brane. 
Here we shall not perform the explicit evaluation of the Euclidean
action in this case. Instead, we use  the analogy with the brane
configuration found in Sec.\ \ref{nuno} 
to evaluate the Euclidean action
naively. 

Since the defect remaining inside the throat is expected to be a
D0-brane, thus the total action in the thin wall approximation 
would be 
\begin{eqnarray}
S_{\rm E}  = -V(S^2) R^3 \epsilon + A(S^2) R^2 S_1 + 2 R T_{{\rm D}0} ,
\end{eqnarray}
here the last term is coming from the D0-brane. 
Then solving the stationary condition for $R$, we obtain
\begin{eqnarray}
 R = \frac{S_1}{\epsilon}
+ \sqrt{
\frac{S_1}{\epsilon} + \frac{T_{{\rm D}0}}{2\pi\epsilon}}.
\end{eqnarray}
Using the previous result
\begin{eqnarray}
 \frac{S_1}{\epsilon} = \frac{2\sqrt{2}c_0}{\sqrt{e}} 
{\alpha'} A = \frac{\sqrt{2}c_0}{\sqrt{e}\pi} l,
\end{eqnarray}
we obtain the Euclidean action as 
\begin{eqnarray}
 S_{\rm E} = \frac{16\pi}{3}\frac{(S_1)^3}{\epsilon^2} + \Delta S
\end{eqnarray}
where
\begin{eqnarray}
 \Delta S = \frac{16 \sqrt{2}c_0}{\sqrt{e}} l_s^2\; l\; T_{{\rm D}2}.
\end{eqnarray}
The ratio of the decay width is given by $\exp(-\Delta S)$, and we
observe that it gets difficult to decay when the D0-brane is left,
and its ratio is almost the exponential of minus of $l/l_s g$. 
Our computation here is using a plausible but conjectural approximation,
and precise  computation may need the gauge field on the
D2-branes. However we believe that our computation is qualitatively
correct.


\subsection{Decay of D0-$\overline{\mbox{D0}}$}

Our computation of the decay width of the D$p$-brane
and the anti D$p$-brane is valid only for $p\geq 1$, since we have
utilized nontrivial bounce solutions in field theories. For the case
$p=0$, that is, a pair of a D0-brane and an anti-D0-brane, then we
have no spatial worldvolume, thus we have to use the quantum
mechanics to describe the annihilation of these D0-branes.

The bounce solution is just a virtual 
particle whose position is described by
$\Phi$ and which rolls down the hill of $-U(\Phi)$ in one dimensional 
space. From the form of the potential we have a bouncing point $\Phi_b$, 
and the Euclidean action is evaluated as 
\begin{eqnarray}
 S_{\rm E} = 2 \int_0^{\Phi_b} d\Phi 
\sqrt{2\left(U(\Phi)-U(0)\right)},
\end{eqnarray}
where $\Phi_b$ satisfies $U(\Phi_b)=U(0)$.
To give the explicit computation we rewrite this expression in terms of 
$T$ as
\begin{eqnarray}
 S_{\rm E} = 
8\sqrt{2} T_{{\rm D}0}
\int_0^{T_b} dT e^{-T^2/2} \sqrt{e^{-T^2}(1+2{\alpha'} A^2 T^2)-1}
\end{eqnarray}
where $T_b$ is the corresponding bounce point.
We shall give here only the numerical result. For a given string
coupling  constant $g=0.1$ and in the unit $l_s =1$, The bouncing point
and the decay width are shown in the following table.
\begin{center}
\begin{tabular}{|c||c|c|c|c|c|c|c|c|c|c|c|c|c|c|c|c|c|}
\hline
$l$ &20 &10 &5 &$\sqrt{2}\pi$ &2 \\\hline
$T_b$ &2.129 &1.638 &0.6748 & 0 & ---  \\\hline
$S_{\rm E}$ &2.1022 &0.8450 &0.06761 &0 & --- \\\hline
\end{tabular}
\end{center}
Note that below the critical separation $l=\sqrt{2}\pi l_s$, there
appears a tachyonic mode at the false vacuum, thus the tunneling
computation becomes invalid in the region $l<\sqrt{2}\pi l_s$. 

We point out here that there exists another interesting decay mode which
is a pair {\it creation} of another D0-brane anti-D0-brane
pair. This pair may be created since the original D0-brane anti-D0-brane
pair produces a background RR gauge field configuration between them as
\begin{eqnarray}
 C_0(x) = 60 \pi^3 g 
\left(
\left(\frac{l_s}{\frac{l}{2}-x}\right)^7
-\left(\frac{l_s}{\frac{l}{2}+x}\right)^7
\right).
\label{back}
\end{eqnarray}
Here we describe the RR gauge potential only along the straight 
line connecting
these two background D0-branes, since only this information is necessary
for the 
later purpose. The coordinate $x$ specifies the point 
on the line, and we have chosen its
zero at the middle point of these background D0-branes.
As we have seen in Sec.\ 2, in the constant RR ($p+2$)-form 
field strength the nucleation of a spherical D$p$-brane may occur. 
Now we are not in the constant field strength but in a varying
background RR field strength, thus a pair creation via non-trivial
(non-spherical)
bounce configuration is expected. In fact, the curvature of the
spherical D-branes is determined only by the background value of the RR
field strength, therefore in this case of the D0-branes a non-trivial
worldline in the Euclidean space may be determined by the RR
configuration (\ref{back}). 

Since the potential (\ref{back}) is strong near the background
D0-branes, we expect that the curvature of the Euclidean D0-brane
worldline is large there, while around the middle point $x=0$ the
curvature becomes small. To obtain a closed line for the bounce solution
we find easily that the consistent bounce configuration must be lie in
the space spanned by $x$ and the Euclidean time $\tau$.  Therefore, we
find that the resultant closed worldline is an ellipse in the $x$-$\tau$
space.  

\begin{figure}[tdp]
\begin{center}
\begin{minipage}{100mm}
\begin{center}
   \leavevmode 
   \epsfxsize=70mm
\put(97,130){$x$}
\put(180,70){$\tau$}
   \epsfbox{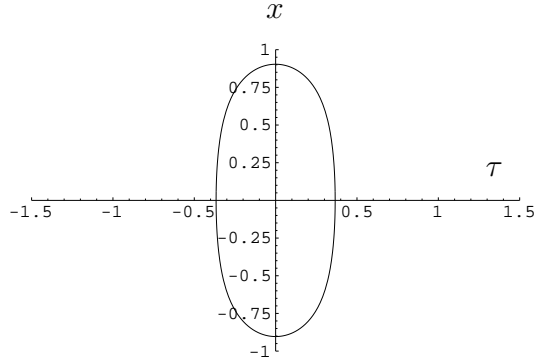}
   \caption{A numerical solution for the bounce with $l=5$. The time
 slice of $\tau = 0$ gives two positions at which a pair of D0-branes
 is created.}
   \label{fig14}
\end{center}
\end{minipage}
\end{center}
\end{figure}

Let us evaluate the decay width by this pair creation process.
The Euclidean action is given by
\begin{eqnarray}
 S = T_{\rm D0}
\int d\tau \left( \sqrt{1+(\p_\tau x)^2} + C_0(x)\right),
\end{eqnarray}
and we can solve this equations of motion numerically. One finds that
the shape of the configuration is actually an ellipse, as shown in Fig.\ 
\ref{fig14}. For $g=0.1$, the numerical results are summarized in the
following table. 
\begin{center}
\begin{tabular}{|c||c|c|c|c|c|c|c|c|c|c|c|c|c|c|c|c|c|}
\hline
$l$ &20 &10 &5 &3 &2 \\\hline
$\Delta l/2$ &8.402 &3.402 &0.9034 &0.0672 &0.0027 \\\hline
$\Delta T/2$ &0.3997 &0.3991 &0.366 &0.0657 &0.0027 \\\hline
$S_{\rm E}$ &409.573 &209.569 &108.184 &18.5413 &0.76387 \\\hline
\end{tabular}
\end{center}
In the table, $\Delta l$ denotes the length of the major axis of the
ellipse while $\Delta T$ denotes that of the minor axis. One observes
that for a large separation $l$ the length of the minor axis does not
grow, and the action becomes proportional to $l$. This means that for a
large separation the nucleation of another pair of D0-branes happens
close to the original positions of the D0-branes, that is, a new
anti-D0-brane is created near the original D-brane and a new D0-brane is
created near the original anti-D0-brane. 
On the other hand, for small separation $l$, it decays very fast and the
shape of the ellipse is almost circular. The spacetime variation of the
background RR field may be neglected in this region. 
After the nucleation of the pair, each nucleated brane will be
approaching to the original D0-brane with opposite charge, 
then we experience a double pair-annihilation. 

After all, compared with the decay width of the naive pair annihilation
in the former analysis, we find that the decay width of the 
another decay mode of the pair creation is small, and it is not a
dominant decay process.


\section{Conclusion}

In this paper, we have studied decay of two typical unstable brane
systems in string theory, from the viewpoint of open string
excitations. The treated instability is due to the quantum 
tunnel effect, a transition from a false vacuum to the true vacuum. 

The first system is the dielectric D2-branes in
the constant RR 4-form field strength (Sec.\ 2). This can decay via
the nucleation of a larger spherical brane which is classically
unstable. We have given explicitly how to compute the decay width (Sec.\
2.4). The new coordinate choice of the worldvolume has enabled us to
perform the computation explicitly. We have found that the tunnel effect
is given by the Euclidean bounce solution which is the funnel-shaped
brane configuration. Another new brane configuration which is a
doughnut-shaped has been found (Sec.\ 2.5),  
and this is relevant for nucleation of a dielectric
brane from nothing. How the nucleated large spherical brane expands has
been also described (Sec.\ 2.3). 

The second system is the brane-antibrane located in parallel with
separation $l$ (Sec.\ 3). 
We have given the action of tachyon field theory on the
brane-antibrane with the parameter $l$ (Sec.\ 3.3). 
BSFT corrections to the tachyon potential has been studied (App.\ A). 
We have observed that this system
sits at the false vacuum, the local minimum of the tachyon potential,
$T=0$.  Applying the generalized Coleman's method, we have obtained the
decay width of the brane-antibrane (Sec.\ 3.4). 
The decay starts with the nucleation
of a true vacuum bubble which is a throat connecting the brane and the
antibrane. We have shown that  
this interpretation is supported by examination of the scalar field
configuration (Sec.\ 3.5).  
The nucleated throat expands accordingly and sweeps out
the brane-antibrane. In addition, we have analysed the tachyon
condensation with leaving topological defects which are stable
lower-dimensional D-branes (Sec.\ 3.6). 
This may be relevant for the braneworld
scenario as discussed later. We have also obtained the decay width of 
D0-$\overline{{\rm D0}}$ in two ways: by the pair annihilation and by
the nucleation of another pair, and found that the former is dominant
decay mode (Sec.\ 3.7). 
The noncommutative generalization has been also studied (App.\ B). 

We have neglected the gravitational effect in this paper. This
approximation may survive the gravitational correction only in limited
situation as in Fig.\ \ref{fig1-1}. String theory necessarily contains
gravity, thus brane physics especially in the large distance scale
should be
influenced by these closed string modes. It might be natural to consider 
a situation in which a brane and an antibrane are created at some time
with separation, and due to the gravitational force and the RR gauge
field exchange the branes attract each other. In this case, as the
branes approach, the form of the tachyon potential varies. At some
distance the quantum tunneling occurs and the throat is nucleated. This
average distance of nucleation of the throat can be estimated by using
our results of the decay width $\Gamma(l)$,  
and the result is roughly a string scale where the true perturbative
instability appears.

One may think that more natural situation would be intersecting
branes. Generally on the intersection the tachyonic mode appears, and
the joining-splitting of the brane surfaces may happen as a result of
the tachyon condensation. We believe that our description of the tachyon
condensation can be applied also on this intersection of the branes. 
The braneworld inflation scenario with the intersecting branes has been
studied recently \cite{inter}.

Since the decay width which we have computed is per the unit time and
volume, severals throat (or spherical branes) may be nucleated
simultaneously. Then the 
bubbles may collide with each other and form a new large bubble. If the
bubbles have charges of the lower dimensional branes, then those branes
may be left inside the large bubble. This may be another mechanism for
leaving lower dimensional branes in the dynamics of the tachyon
condensation. 

While the collision of the throats, most of the energy of the moving
domain walls may be converted to the energy of the radiation of the
closed string modes in the bulk. Other than this, there are massive open
string modes which may propagate in the false vacuum. We employed the
effective field theory of the tachyon field, thus it receives ${\alpha'}$
corrections which we neglected in this paper. Inclusion of open string
massive modes and also closed string modes such as gravity is important.

Considering the closed string massless modes amounts to the inclusion
of the back reaction which we have not considered in this paper. 
On the other hand, another interesting viewpoint is that there may be 
the gravity description which is ``dual'' to our open string
description,\footnote{We would like to thank S.\ Hirano for pointing out 
this viewpoint.} as suggested from the $s\leftrightarrow t$ channel
duality 
of the string worldsheet theory and the AdS/CFT correspondence. It is
intriguing to study some duality between ours and the gravitational
instabilities studied so far 
(see for example \cite{grain} for the gravitational description
of the instabilities of some backgrounds). 
We leave these  for the future work.

\vspace{20mm}


\appendix

\noindent
{\Large {\bf  Appendix}}

\vspace{-7mm}

\section{BSFT corrections to the tachyon potential}

To obtain a more precise expression for the tachyon potential $V(T)$,
let us study the corrections of higher orders in $l$ derived from the
BSFT.  
As seen above, the $l$-dependence of the potential is coming from
the covariant derivative through the T-duality. Therefore, the first
nontrivial corrections might come from the results of linear profiles in 
brane-antibrane system in the BSFT \cite{Terashima}.
Their result shows that, if we turn on a part of the tachyon which has
a linear profile only along a certain direction in the target space, 
the BSFT action is
\begin{eqnarray}
 V(T) = e^{-|T|^2}F(2 {\alpha'}|\p_iT|^2)
\label{A1}
\end{eqnarray}
where
\begin{eqnarray}
 F(x) \equiv \frac{4^x x \Gamma(x)^2}{2\Gamma(2x)}.
\end{eqnarray}
Since we turn on only the constant Wilson line, the gauge field strength 
vanishes, thus the dependence on the gauge field comes only from the
covariant derivative. We lift the above expression to include  the 
covariant derivative by simply changing the argument of the function 
$F$. Noting that the T-duality is just a dimensional reduction, 
we obtain the resultant expression for the corrected tachyon potential
for the brane-antibrane system :
\begin{eqnarray}
 V(T) = e^{- |T|^2}F(2{\alpha'}A^2|T|^2).
\label{potenbsft}
\end{eqnarray}

In the above argument we have used the linear tachyon profile.
One may wonder if the $l$ dependence of the tachyon potential 
comes also from the higher derivative corrections such as 
$(DDT)^2 \sim l^4 T^2$. 
We can derive exact tachyon potential valid for all order in
$l$ as in the following. Let us recall how the BSFT action is obtained.  
First we have the boundary interaction of a string sigma model
\cite{Terashima} 
\begin{eqnarray}
 S_{\rm bdry} = -\int_{\rm boundary} d\tau
\left[
-\frac{{\alpha'}}{4}T^I T^I+\frac14 \eta^I \dot{\eta}^I 
+ \frac{{\alpha'}}{2}D_\mu T^I \psi^\mu \eta^I
\right].
\end{eqnarray}
We have written only the terms which are relevant for our calculation,
and $I=1,2$, $T=(T^1+iT^2)/\sqrt{8\pi{\alpha'}}.$
After integrating out $\eta$ and adding the worldsheet action, we obtain
\begin{eqnarray}
 S_{\rm worldsheet} + S_{\rm bdry} =
|T|^2 + \frac12 \sum_{n=1}^{\infty}n X_{-n} X_n
+ i \sum_{r=1/2}^{\infty}
\left(1+ \frac{2{\alpha'}A^2|T|^2}{r}\right)\psi_{-r} \psi_r.
\end{eqnarray}
Therefore the $A$ dependence in the partition function is given by
\begin{eqnarray}
 Z \propto 
\frac{\prod_{r=1/2}^{\infty} 
\left(1+\frac{2{\alpha'} A^2 |T|^2}{r}\right) }
{\prod_{n=1}^{\infty} n }.
\end{eqnarray}
Finally, by restoring the appropriate normalization, we have 
\begin{eqnarray}
 Z 
=2T_{\rm D3}\int d^4X_0 e^{-2\pi{\alpha'} |T|^2}
\frac{4^x \Gamma(x)}{2 \Gamma(2x)},
\label{potbsft}
\end{eqnarray}
where $x=2{\alpha'} A^2 |T|^2$.
Note that this expression is a bit different from the partition function
for the brane-antibrane system for the linear tachyon profile (\ref{A1}).

\begin{figure}[tdp]
\begin{center}
\begin{minipage}{100mm}
 \begin{center}
   \leavevmode
   \epsfxsize=70mm
\put(20,120){$V(T)$}
\put(180,20){$T$}
   \epsfbox{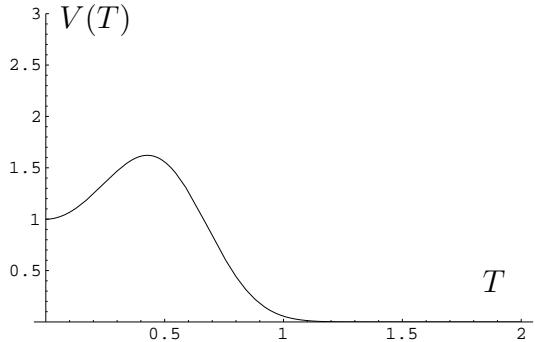}
   \caption{the potential from BSFT with the same values of the
  parameters as in Fig.\ \ref{fig1}. }
   \label{fig2}
  \end{center}
\end{minipage}
\end{center}
\end{figure}

The final potential 
is shown in Fig.\ \ref{fig2}. We can see that the form of the
potential is qualitatively the same. One should note that the Taylor
expansion of the expression (\ref{potbsft}) in terms of $A^2$ does not
match (\ref{potential}), because the BSFT result does depend on the
renormalization scheme which is unknown in this context. The difference
of the scale of these figures might come from that finite
renormalization. 
One of the reasons why for the computation of the decay width we have
used (\ref{potential}) instead of the  exact BSFT result (\ref{potbsft})
is that,  the BSFT result should include also the higher order
derivative coupling of the tachyon fields, and it is very hard to adopt
this for a practical computation. The potential (\ref{potential}) was
derived from the two derivative truncation, thus for the analysis with
the kinetic terms of the two derivative truncation it might be better to
use the tachyon potential derived in the same truncation. 
We believe that the full potential after the finite
renormalization may have a similar structure.

\section{Noncommutative solitons}

One of the early attempts to understand the decay 
of the brane-antibrane system is found in \cite{awata}. 
The authors of \cite{awata} tried to construct the brane-antibrane
effective action by using Matrix theory. Now we understand that
this construction leads to a noncommutative worldvolume.\footnote{For
  relevant discussion on the throat solution of the scalar field 
  under the constant electro-magnetic field, see \cite{koji}.}

On the noncommutative worldvolume one can construct nontrivial
co-dimension two solitons which are called noncommutative solitons
\cite{GMS}. The noncommutative scalar solitons found their intriguing
application in the tachyon condensation in string theory
\cite{com,harvey}.  
Since in the text  we have not obtained analytic form of
the bounce solution, the introduction of the noncommutativity may
help us to construct it explicitly. 
In this appendix, 
we apply this method to our potential, and obtain stable solitons
which are analogues of \cite{GMS}. We give the brane interpretation of
them.  

First, let us briefly review what was found in \cite{GMS}. 
After an appropriate scaling of the background metric,
B-field and ${\alpha'}$, we obtain noncommutative worldvolume theory of
the brane-antibrane. We assume that the result is simply the replacement 
of the ordinary field multiplication with the star product defined with
the noncommutativity parameter $\theta$. 
In the large noncommutativity limit, the authors of \cite{GMS}
constructed a soliton solution whose form is determined only by the
information of the extrema of the scalar potential:
\begin{eqnarray}
  \Phi_{\rm sol} = \sum_i \lambda_i P_i.
\label{ncsol}
\end{eqnarray}
Here $\lambda_i$ is the value of $\Phi$ which makes the potential
extrema, and $P_i$ is the projection operator which satisfies $P_i=P_i
* P_i$. Note that as explicitly written in \cite{com} the potential
should have its local extrema at $\Phi=0$. 

One can easily find that our potential written by $\Phi$ satisfies
this requirement, and we have two extrema. One of them is the true
vacuum $\Phi = \Phi_1 = \sqrt{4\pi{\alpha'} T_{{\rm D}p}}$. 
The stability argument
developed in \cite{GMS} shows that if we choose $\Phi_1$ as
$\lambda$, then the solution (\ref{ncsol}) is classically stable.

Let us consider what is this stable brane configuration. The simplest
choice for the projection operator is 
\begin{eqnarray}
  P^{(1)} = 1 - |0\rangle\langle 0 | 
\quad \mbox{or} \quad
  P^{(2)} = |0\rangle\langle 0 |.
\end{eqnarray}
It is easy to consider $P^{(1)}= 1-P^{(2)}$ first.
Note that the above $P^{(2)}$ written in an operator
representation is identical with a Gaussian function whose width is
given by $\sim \sqrt{\theta}$. Thus the solution $\Phi_{\rm sol} =
\Phi_1 P^{(1)}$ is almost sitting at the true vacuum except for the
disc region whose radius is $\sqrt{\theta}$ and whose center is at the
origin of the brane-antibrane worldvolume. 
Inside this disc, the tachyon is not condensed and thus the brane
worldvolume exists, while outside of the disc the brane worldvolumes
vanish. Since two worldvolumes are separated now, one may imagine
that the resultant form of the brane is spherical as if it were
a dielectric brane. However, there is no mechanism to pop up branes
in the constant B-field background (if one turns on a constant
H-field, then one can have a popped-up brane \cite{myers}). 
Then, what is this brane configuration?

The resolution of this puzzle is easily found if one notice that
the ``size'' $\sqrt{\theta}$ of the disc region is a fake, since in
the noncommutative worldvolume it is impossible to give a notion of
precise location, because of the noncommutativity. One has to move to
the commutative description via the Seiberg-Witten map
\cite{SWmap, tachyonSW}. 
Although the SW map for the tachyon field has not been
obtained (see for example \cite{tachyonSW}), 
from the discussion in \cite{HO},
the natural result of the SW map for the above noncommutative soliton 
is a delta function. Thus, the actual size of the above discs is
zero. This gives an interpretation of the above soliton. A size zero
D0-brane and an anti-D0-brane are located at the centers of the 
original D2-branes, respectively, with the separation $l$. 
This interpretation is
consistent with the observation in \cite{NCsol}. 
Now the separation length
between the D0-brane and the anti-D0 is enough large, this system is
classically stable. 

If one employs another projection operator $P^{(2)}$, then easily
the brane interpretation of them can be given : Each D2-brane has a tiny 
hole at each center. Thus this classically stable solution is singular
and is not relevant for the decay of the brane-antibrane. 

\vs{10mm}
\noindent
{\large \bf Acknowledgements}

The author is grateful to S.\ Nagaoka for valuable discussions,
and would like to thank H.\ Hata, C.\ Herdeiro, 
S.\ Hirano, G.\ Horowitz, 
N.\ Sasakura and T.\ Takayanagi for useful comments.

\newcommand{\J}[4]{{\sl #1} {\bf #2} (#3) #4}
\newcommand{\andJ}[3]{{\bf #1} (#2) #3}
\newcommand{\AP}{Ann.\ Phys.\ (N.Y.)}
\newcommand{\MPL}{Mod.\ Phys.\ Lett.}
\newcommand{\NP}{Nucl.\ Phys.}
\newcommand{\PL}{Phys.\ Lett.}
\newcommand{\PR}{ Phys.\ Rev.}
\newcommand{\PRL}{Phys.\ Rev.\ Lett.}
\newcommand{\PTP}{Prog.\ Theor.\ Phys.}
\newcommand{\hep}[1]{{\tt hep-th/{#1}}}

\end{document}